\documentclass[a4paper,11pt]{article}
\pdfoutput=1 
\RequirePackage{snapshot}  
\usepackage{jheppub} 
\usepackage{bm,amsmath,amssymb,slashed,graphicx,%
            enumerate,alltt,xspace,multirow,xcolor,mathrsfs}
\usepackage{fancyvrb}
\usepackage{booktabs,tabularx}
\usepackage{graphicx}
\usepackage{subcaption}
\usepackage{xspace}
\usepackage{cancel}
\usepackage[utf8]{inputenc} 
\usepackage[compat=1.0.0]{tikz-feynman}
\usepackage{scalerel}
\usepackage{layouts}
\usepackage{adjustbox}
\usepackage{pdflscape}
\usepackage[normalem]{ulem} 
\usepackage{lineno}
\usepackage{wasysym}

\usepackage{algorithm}
\usepackage[noEnd=true]{algpseudocodex}

\newcommand{\dd}{\mathrm{d}}

\newcommand{\ktcut}{k_{t,\mathrm{cut}}}
\newcommand{\DL}{\text{DL}}
\newcommand{\NDL}{\text{NDL}}
\newcommand{\NNDL}{\text{NNDL}}

\newcommand\betaps{\beta_\text{\textsc{ps}}}

\usepackage[capitalise]{cleveref}

\makeatletter
\g@addto@macro\bfseries{\boldmath}
\makeatother

\definecolor{labelkey}{rgb}{0,0.5,0.0}
\definecolor{royalpurple}{rgb}{0.47, 0.32, 0.66}

\usepackage{listings}
\definecolor{darkred}{rgb}{0.6,0.0,0}
\definecolor{darkgreen}{rgb}{0,0.4,0}

\definecolor{grey}{rgb}{0.5,0.5,0.5}
\definecolor{rust}{rgb}{0.9,0.4,0.0}
\definecolor{lightblue}{rgb}{0.0,0.5,1.0}

\allowdisplaybreaks 

\definecolor{semiblue}{rgb}{0.3,0.3,0.8}
\newcommand{\logbook}[2]{}

\newcommand{\codefont}[1]{\texttt{#1}}

\newcommand{\NODS}{\codefont{NODS}\xspace}    
\newcommand\nods{\NODS}
\newcommand{\pythia}{\codefont{Pythia}\xspace}

\newcommand\pythiaeight{{\pythia$\,$8}\xspace}

\newcommand{\pg}{{PanGlobal}\xspace}
\newcommand{\pl}{{PanLocal}\xspace}
\newcommand{\panscales}{{PanScales}\xspace}
\newcommand{\dire}{{Dire-v1}\xspace}

\newcommand{\as}{\alpha_s}

\newcommand{\dipm}{\widetilde{m}_{ij}}

\newcommand{\itilde}{{\widetilde \imath}}
\newcommand{\jtilde}{{\widetilde \jmath}}

\newcommand\sss{\mathchoice%
{\displaystyle}%
{\scriptstyle}%
{\scriptscriptstyle}%
{\scriptscriptstyle}%
}

\newcommand{\rl}{r_{\rm \sss L}}

\PassOptionsToPackage{hypertexnames=false}{jheppub}

\tikzstyle{block} = [rectangle, minimum width=1.0cm, minimum height=0.75cm, thin, draw=black]
\tikzstyle{blob} = [circle, minimum width=0.5cm, thin, draw=black]
\tikzset{blackarrow/.style={-stealth, semithick, draw=black}}
\tikzset{connection/.style={inner sep=0,outer sep=0}}

\newcolumntype{C}{>{\centering\arraybackslash}X}

\title{Logarithmically-accurate final-state showers with massive quarks}

\preprint{\small Nikhef 2026-007 }

\newcommand{\Torinoaff}{Dipartimento di Fisica, Università di Torino, and INFN, Sezione di Torino,
Via P. Giuria 1, I-10125 Torino, Italy}
\newcommand{\NIKHEFaff}{Nikhef, Theory Group, Science Park 105, 1098 XG, Amsterdam, The Netherlands}
\newcommand{\CNRSaff}{Université Paris-Saclay, CNRS, CEA, Institut de physique théorique, 91191, Gif-sur-Yvette, France}
\newcommand{\UGRaff}{Departamento de F\'{i}sica Te\'{o}rica y del Cosmos, Universidad de Granada,
Campus de Fuentenueva, E-18071 Granada, Spain}
\newcommand{\UCLaff}{Department of Physics and Astronomy, University College London, London, WC1E 6BT, UK}

\author[a]{Melissa van Beekveld,}%
\author[b]{Silvia Ferrario Ravasio,}%
\author[c]{Alba Soto-Ontoso,}%
\author[d]{Gregory Soyez,}%
\author[e]{Rob Verheyen}%

\affiliation[a]{\NIKHEFaff}
\affiliation[b]{\Torinoaff}
\affiliation[c]{\UGRaff}
\affiliation[d]{\CNRSaff}
\affiliation[e]{\UCLaff}

\date{Received: date / Accepted: \today}

\abstract{
We formulate \panscales final-state showers that account for quark masses and achieve next-to-leading logarithmic accuracy, while preserving the original accuracy of the showers for observables where the mass of the quarks is irrelevant.
We validate the accuracy of the shower algorithms by performing fixed-order tests up to second order in the strong coupling constant, and all-order comparisons to (semi-)analytic resummed calculations for a series of observables, including Lund-tree shapes, non-global energy flows and Lund sub-jet multiplicities. 
We also include some phenomenological studies using LEP data.
}

\keywords{QCD, Parton Shower, Masses, Resummation, LHC, LEP
  \\[4em]
  \textit{For the purpose of Open Access, the authors have applied a CC BY
  public copyright licence to any Author Accepted Manuscript (AAM)
  version arising from this submission.}
}

\begin{document}
\maketitle

\section{Introduction}

Heavy quarks such as bottom and charm are produced abundantly in collider experiments such as the Large Hadron Collider (LHC). 
Their physics is intimately connected with the study of the Higgs boson and the top quark and therefore of great interest for advancing our understanding of the fundamental interactions.
From a phenomenological perspective, the advance of jet flavour tagging techniques~\cite{Caletti:2022hnc,Czakon:2022wam, Gauld:2022lem, Caola:2023wpj, Behring:2025ilo} demands an improvement of the modelling of QCD radiation in jets containing massive partons. 

The radiation pattern of QCD jets involving heavy-quarks displays a series of distinct features compared to the massless case. 
The most prominent one is the suppression of collinear emissions within a cone of parametric size $m_Q/E$ around the heavy quark direction, with $m_Q$ and $E$ the heavy quark mass and energy, an effect known as the dead-cone~\cite{Dokshitzer:1991fd}. 
The experimental quest to unveil this fundamental property of QCD started with the LEP programme. 
This high-energy collider studied final-states involving heavy-quarks and measured fragmentation functions~\cite{OPAL:1995rqo,SLD:1999cuj,ALEPH:2001pfo} and charged-particle multiplicities~\cite{SLD:1996yvs,DELPHI:2000edu}. 
Heavy-quark initiated process displayed a harder fragmentation pattern and higher multiplicities with respect to the light-quark case. The harder fragmentation pattern follows from the radiation pattern from heavy quarks, which retain a larger fraction of their original energy than light quarks, whereas the observed higher multiplicities are mainly related to the contribution of heavy-hadron decay products.
However, neither observation provided direct access to the dead-cone.
More recently, the LHC experiments have achieved the first observation of the dead cone by using jet substructure techniques~\cite{Cunqueiro:2018jbh} in charm-tagged jets~\cite{ALICE:2021aqk}. 
The heavy-flavour program at the LHC has also released a suite of measurements of $b$-quark fragmentation functions~\cite{ATLAS:2021agf,LHCb:2025zmu} and jet substructure observables~\cite{CMS:2018ypj,ATLAS:2022miz,ALICE:2022phr,ALICE:2025igw,LHCb:2025tvf,LHCb:2025mcq}. 
These datasets are crucial to test our understanding of QCD radiation in the presence of masses, and to constrain the modelling of heavy-flavour jets in parton-shower Monte Carlo event generators. 

From the perturbative QCD point of view, the presence of a mass scale in the problem leads to the appearance of new logarithmically-enhanced terms in the perturbative expansion, which can be large when the mass is small compared to the hard scale of the process. For instance, in the case of a heavy-quark fragmentation function, the resummation of these logarithms has been performed at next-to-leading logarithmic (NLL) accuracy in Refs.~\cite{Mele:1990cw, Cacciari:2001cw,Cacciari:2002xb}, and more recently at next-to-next-to-leading logarithmic (NNLL) accuracy in Ref.~\cite{Maltoni:2022bpy,Bonino:2023icn}.
The resummation of mass logarithms for LHC observables is an active topic of research~\cite{Craft:2022kdo,Caletti:2023spr,Dhani:2024gtx,Ghira:2025nym,Barata:2025uxp, Liu:2025ldi}, and a general framework to perform such resummation for generic observables is still missing. 

Parton showers are a powerful tool to achieve this goal, as they can
be designed to capture the logarithmic structure of QCD radiation in
the presence of masses. At first, mass effects were incorporated into
parton showers at LL accuracy
~\cite{Norrbin:2000uu,Gehrmann-DeRidder:2011gkt}. More recently, NLL
accuracy for global event shapes was achieved in the angular-ordered
shower in Herwig both in the massless~\cite{Bewick:2019rbu} and
massive cases~\cite{Hoang:2018zrp}. Alaric has demonstrated NLL
accuracy for global event shapes when the mass is irrelevant (i.e.\
below the scale of the observable)~\cite{Herren:2022jej}, but not
explicitly in the massive case~\cite{Assi:2023rbu}.
Validating the logarithmic accuracy of a parton shower in the presence of quark masses requires extending the procedure first introduced in Ref.~\cite{Dasgupta:2020fwr}, and used in Refs.~\cite{Hamilton:2020rcu,Karlberg:2021kwr,Hamilton:2021dyz,vanBeekveld:2022ukn,Herren:2022jej,Hamilton:2023dwb,FerrarioRavasio:2023kyg,vanBeekveld:2024wws,Preuss:2024vyu,vanBeekveld:2025lpz,Helliwell:2025krf}. Namely, one needs to test whether the shower correctly captures logarithms of the mass, as well as logarithms of the infrared scale. 

In this work we introduce \panscales showers that account for quark masses and we validate their NLL accuracy. 
More generally, our goal is to provide a framework that systematically captures the logarithmic structure of mass effects across different classes of observables. 
In practice, there are two specific kinematic limits in which
heavy-quark effects should be taken into account.
The first is the {\it quasi-collinear limit}, i.e.\ the limit where
the angle between two massive final-state particles is small and of the same
order as the dead-cone size,\footnote{If these two particles have
  masses $m_i, m_k$, momenta $p_i, p_k$, and a relative transverse
  momentum $k_{\perp}$, this corresponds to taking the limit $p_i
  \cdot p_k \sim m_i^2, m_k^2, k_{\perp}^2 \to 0$.} where
a correct treatment of the massive splitting functions is critical for heavy-quark fragmentation functions.
The second limit corresponds to the radiation of soft gluons at
angles commensurate with the dead-cone size (small or not) where the
mass affects the structure of non-global
logarithms~\cite{Dasgupta:2001sh, Banfi:2002hw,Balsiger:2020ogy} typically for
observables restricted to a limited angular region. 
This is for example the case for the jet masses and hemisphere
observables used in top-mass extractions.
The results are publicly available, and interfaced to \pythiaeight~\cite{Bierlich:2022pfr} for hadronisation, as of the 0.4.0 release of the \panscales code~\cite{vanBeekveld:2023ivn}.
The paper is organised as follows. 
In Sec.~\ref{sec:formulation} we present the formulation of the \panscales family of showers when massive quarks are involved. 
We discuss the kinematic map and emission probability for both the \pg and \pl variants, as well as the implementation of subleading colour and spin correlations. 
In Secs.~\ref{sec:fo-val} and \ref{sec:log-accuracy-tests} we validate the logarithmic accuracy of our showers by performing a series of fixed-order and all-order tests. 
We present some phenomenological studies using LEP data in Sec.~\ref{sec:pheno} and conclude in Sec.~\ref{sec:conclusions}. 
Appendices~\ref{app:mapping-coefficients} and \ref{app:analytics-multiplicity} contain technical details on the kinematic maps, and the resummation of Lund multiplicities, respectively. 
Finally, App.~\ref{app:simple-veto} discusses a simple shower
algorithm with massless kinematics and local recoil which achieves NLL
accuracy for global massive event shapes with a straightforward
modification of standard shower algorithms. 
This prescription should however be
viewed as an illustration of how NLL accuracy can be reached and it is important to stress that this simplified algorithm does not capture non-global logarithms.

\section{\panscales showers with massive quarks}
\label{sec:formulation}
In this section we present the formulation of the
\panscales showers when massive quarks are involved. 
We consider both an antenna-like shower with global transverse-momentum conservation, called \pg, whose massless formulation can be found in Refs.~\cite{Dasgupta:2020fwr,FerrarioRavasio:2023kyg}, and a dipole variant with fully-local momentum conservation, named \pl~\cite{Dasgupta:2020fwr}.
We begin by describing elements that are common to the formulation of both variants.

We consider an emission with momentum $p_k$ radiated from a
(potentially) massive dipole with pre-branching momenta ($\widetilde{p}_i$,
$\widetilde{p}_j$), such that $\widetilde{p}_i^2=\widetilde{m}_i^2$
 and $\widetilde{p}_j^2=\widetilde{m}_j^2$. In the massless formulation of the \panscales showers, the emission’s momentum $p_k$ as well as the post-branching momenta of the dipole legs ($p_i, p_j$) are Sudakov-decomposed in terms of ($\widetilde{p}_i$, $\widetilde{p}_j$) and a transverse component. When masses are involved, we instead use \emph{lightlike vectors}, which we denote as ($\widetilde{n}_i$, $\widetilde{n}_j$), as basis for the Sudakov decomposition. We design the lightlike vectors such that, in the dipole rest frame, their three-momenta are equal to those of the massive dipole legs. Physically, this means that only the energy components of ($\widetilde{n}_i$, $\widetilde{n}_j$) and ($\widetilde{p}_i$, $\widetilde{p}_j$) differ in this rest frame. Their explicit definition is
\begin{subequations}\label{eq:light-like-ref}
\begin{align}
  \widetilde n_i & = \widetilde{p}_i + x_i (\widetilde{p}_i+\widetilde{p}_j)
               = (1+x_i) \widetilde{p}_i + x_i \widetilde{p}_j\,, \\
  \widetilde n_j & = \widetilde{p}_j + x_j (\widetilde{p}_i+\widetilde{p}_j)
               = x_j \widetilde{p}_i + (1+x_j) \widetilde{p}_j\,.
\end{align}
\end{subequations}
The coefficients $x_i$ and $x_j$ are obtained by imposing $\widetilde
n_i^2=\widetilde n_j^2=0$ and read
\begin{subequations}\label{eq:x-coefficients}
\begin{align}
  x_i &
       =\frac{1}{2}\left[\widetilde\kappa_{ij}-(1+\widetilde \mu_i-\widetilde \mu_j)\right] ,\\
  x_j &
       =\frac{1}{2}\left[\widetilde\kappa_{ij}-(1-\widetilde \mu_i+\widetilde \mu_j)\right]\, ,
\end{align}
\end{subequations}
where we have introduced
\begin{align}\label{eq:kallen2}
  \widetilde{\mu}_{i,j} = \frac{\widetilde{m}_{i,j}^2}{\dipm^2}\, ,
  \quad \text{ with } \quad
  \dipm^2 = (\widetilde{p}_i + \widetilde{p}_j)^2\,, 
\end{align}
and
\begin{align}\label{eq:kallen}
  \widetilde\kappa_{ij}^2 = \lambda(1, \widetilde{\mu}_i, \widetilde{\mu}_j)\, ,
  \quad \text{ with } \quad
\lambda(a,b,c) = (a-b-c)^2 - 4 b c\,. 
\end{align}
The choice of lightlike vectors introduced in Eq.~\eqref{eq:light-like-ref} enjoys a number of useful properties. First, the coefficients $x_i$ and $x_j$ become zero in the massless limit. This means that $\widetilde{n}_{i,j}$ is equal to $\widetilde{p}_{i,j}$ when $\widetilde{p}_{i,j}$ is massless, regardless of the mass of the other particle present in the dipole. Secondly, they satisfy
\begin{equation}
\widetilde{n}_i + \widetilde{n}_j = \widetilde{\kappa}_{ij}\left(\widetilde{p}_i + \widetilde{p}_j\right),
\label{eq:sumntilde}
\end{equation}
meaning that their dot product is proportional to the mass of the dipole squared, $2\widetilde{n}_i\cdot \widetilde{n}_j = \widetilde{\kappa}_{ij}^2\dipm^2$. %
We found that preserving the direction makes it easier to meaningfully define the quasi-collinear momentum fraction $z$.
In addition, from a numerical perspective, we found that this choice
of reference vectors improves the stability of logarithmic accuracy tests.

We also use these lightlike vectors to define a series of invariants
\begin{equation}
  \widetilde s_{i}  = 2\widetilde n_{i}\cdot Q, \quad
  \widetilde s_{j}  = 2\widetilde n_{j}\cdot Q, \quad
  \widetilde s_{ij} = 2\widetilde n_{i}\cdot\widetilde n_{j},
\end{equation} 
where $Q$ is the total momentum in the event frame. Note that in the massless version of the \panscales showers, these invariants are defined using $(\widetilde p_i, \widetilde p_j)$. The shower emission probability is then expressed in terms of an evolution variable $v$, a rapidity-like variable $\bar{\eta}$, and an azimuthal angle $\phi$. Out of them, we build an effective transverse momentum
\begin{equation}
  \label{eq:kappatdefn}
  \kappa_t = \left(\frac{\widetilde s_i\widetilde s_j}{Q^2 \widetilde s_{ij}}\right)^{\frac{\beta}{2}} v e^{\beta|\bar\eta|}\,,
\end{equation}
and two auxiliary-variables related to the longitudinal momentum fractions of the emission with respect to each dipole leg
\begin{equation}
  \label{eq:akbkdefn}
  \alpha_k = \sqrt{\frac{\widetilde s_j}{\widetilde s_{ij}\widetilde s_i}}\kappa_t e^{+\bar\eta}, \quad
  \beta_k  = \sqrt{\frac{\widetilde s_i}{\widetilde s_{ij}\widetilde s_j}}\kappa_t e^{-\bar\eta}\,.
\end{equation}
With these definitions, we can write the differential emission probability for a dipole leg $\itilde$ to emit a parton $k$ as
\begin{equation}
  \label{eq:singleleg-prob}
\dd \mathcal{P}_{\itilde(\jtilde)\to ik(\jtilde)} = \dd\bar{\eta}\frac{\dd\phi}{2\pi} \dd\ln v \frac{\partial \ln \kappa_t}{\partial \ln v}\frac{\alpha_s^{\rm eff}(\kappa_t)}{\pi}\alpha_k P_{\itilde \to ik}(\alpha_k),
\end{equation}
where $\jtilde$ is the other dipole leg.
The total differential emission probability for a dipole $(\itilde,\jtilde)$ to emit a parton $k$ is instead defined as
\begin{align}\label{eq:shower-prob}
\dd \mathcal{P}_{\itilde\jtilde\to ijk} = g(\bar{\eta}) \dd \mathcal{P}_{\itilde (\jtilde)\to ik(\jtilde)}  + g(-\bar{\eta})\dd \mathcal{P}_{\jtilde(\itilde)\to jk(\itilde)}. 
\end{align}
Let us discuss each of the ingredients entering this expression. The effective shower coupling, $\alpha_s^{\rm eff}$, is evaluated in a variable-number flavour scheme, as will be discussed in Sec.~\ref{sec:kcmw-threshold}. The function $g(\bar\eta)$ is a partition function that ensures a smooth transition around $\bar \eta=0$ when the role of the emitter changes. Its explicit expression for the \pg and \pl showers can be found in Ref.~\cite{Dasgupta:2020fwr}. Finally, $P_{\itilde \to ik}(\alpha_k)$ and $P_{\jtilde \to jk}(\beta_k)$ are the branching probabilities associated with each dipole leg. To achieve NLL accuracy, one requirement is that the shower must reproduce the exact QCD emission probability when the emission is soft or (quasi-)collinear to either of the momenta that are present in the emitting dipole. That is, Eq.~\eqref{eq:shower-prob} must reduce to 
\begin{equation}
  \label{eq:dP-soft-limit}
  \dd \mathcal{P}_{\itilde\jtilde\to ijk} \overset{\text{soft}}{=} 16\pi\alpha_s T_{i}\cdot T_{j} \left[\frac{\widetilde{p}_i \cdot \widetilde{p}_j}{\widetilde{p}_i \cdot p_k \, \widetilde{p}_j \cdot p_k} - \frac{\widetilde{m}_i^2}{2(\widetilde{p}_i \cdot p_k)^2}- \frac{\widetilde{m}_j^2}{2(\widetilde{p}_j \cdot p_k)^2}\right]\dd\Phi_{\rm soft}\,,
\end{equation}
when the emission $k$ is soft, i.e. when $E_k \ll \widetilde{m}_{ij},\widetilde m_i, \widetilde m_j$. In Eq.~\eqref{eq:dP-soft-limit} we have introduced colour charge operators of SU(3), $T_{i}$.
For gluons, $T_i$ coincides with a SU(3) matrix in the adjoint representation, for (anti-)quarks in the (anti-)fundamental.
The factor in the square bracket reflects a dynamical suppression of radiation at angles smaller than $\widetilde m_i/\widetilde E_i$, or $\widetilde m_j/\widetilde E_j$.

On the other hand, when $i$ and $k$ become quasi-collinear, i.e. when  $p_i\cdot p_k , m^2_i , m^2_j \ll m^2_{ij}$, Eq.~\eqref{eq:shower-prob} must reduce to
\begin{equation}
  \label{eq:dP-col-limit}
  \dd \mathcal{P}_{\itilde\jtilde\to ijk} \overset{\text{col}}{=} 8\pi\alpha_s \frac{1}{p_i\cdot p_k + m_i^2+m_k^2-\widetilde{m}_i^2}P^{\rm AP}_{\itilde\to i k}(z) \dd\Phi_{\rm col} \,.
\end{equation}
The helicity-averaged massive DGLAP splitting functions, $P^{\rm AP}_{\itilde\to i k}(z)$, depend on the longitudinal momentum fraction $z$ of particle $k$. Their explicit expressions for heavy-quark emissions read
\begin{subequations}
\begin{align}
  P^{\rm AP}_{Q\to Qg}(z) &= C_F\left[\frac{1+(1-z)^2}{z}-\frac{m_Q^2}{p_i\cdot p_k}\right],\\
  P^{\rm AP}_{g\to Q\bar Q}(z) &= T_R\left[z^2+(1-z)^2+\frac{m_Q^2}{p_i\cdot p_k+m_Q^2}\right].
\end{align}
\end{subequations}
In Eqs.~\eqref{eq:dP-soft-limit} and~\eqref{eq:dP-col-limit}, $\dd\Phi_{\rm soft}$ and $\dd\Phi_{\rm col}$ correspond to the appropriate limit of the antenna phase-space element which reads
\begin{equation}
  \dd\Phi = \frac{1}{4 \pi^2}
  \frac{1}{\dipm^2\widetilde\kappa_{ij}} \dd(p_i\cdot p_k)\dd(p_j\cdot p_k) \frac{\dd\varphi}{2\pi}.
\label{eq:antennaphsp}
\end{equation}
In Secs.~\ref{sec:pl-def} and~\ref{sec:pg-def} below, we discuss how
we construct the kinematic map and emission probability for the \pl
and \pg showers so that they reproduce these limits.

We account for subleading colour effects using the nested ordered double-soft (\nods) method introduced in Ref.~\cite{Hamilton:2020rcu}. We have not modified the \nods algorithm compared to the massless case, which means that subleading colour corrections are not fully taken into account in the dead-cone region. Quasi-collinear spin-correlations are included by adapting the Collins-Knowles algorithm~\cite{Collins:1987cp,Knowles:1988hu} as implemented and extended in Refs.~\cite{Karlberg:2021kwr,Hamilton:2021dyz} to account for mass corrections. More details on the spin implementation are presented in Sec.~\ref{sec:spin-correlations}.  

\subsection{\pl}
\label{sec:pl-def}
\subsubsection{Kinematic map}
The \pl shower implements dipole-local recoil, with the transverse momentum imbalance fully absorbed by the emitting particle, whereas the spectator particle only absorbs some longitudinal recoil.
The post-branching momenta are written in terms of the pre-branching lightlike vectors as
\begin{subequations}\label{eq:pl-map}
\begin{align}
    p_k^\mu &= \alpha_k \widetilde n_i^\mu + \beta_k \widetilde n_j^\mu - k_{\perp}^\mu \,, \\
  p_i^\mu &= a_i \widetilde n_i^\mu + b_i \widetilde n_j^\mu + k_{\perp}^\mu\, ,\\
  p_j^\mu &= a_j \widetilde n_i^\mu + b_j \widetilde n_j^\mu\, ,
\end{align}
\end{subequations}
with
\begin{align}
k_{\perp}^\mu = k_t \hat{n}_\perp^\mu, \qquad \hat n_\perp^{\mu} = \hat n_{\perp,1}^\mu \cos\phi + \hat n_{\perp,2}^\mu \sin\phi\,,\qquad \hat n_{\perp,l}^2 = -1\,,\qquad \hat n_{\perp,l}\cdot \widetilde{n}_{i,j} = 0\,,
\label{eq:nperp}
\end{align}
and
\begin{equation}
k_t^2 =2 \alpha_k \beta_k \widetilde{n}_i \cdot \widetilde{n}_j -m_k^2,
\label{eq:kt2fornperp}
\end{equation}
with $\alpha_k$ and $\beta_k$ defined as in Eq.~\eqref{eq:akbkdefn}.
We can solve for the mapping coefficients $a_i,a_j,b_i,b_j$ by
imposing momentum conservation along the longitudinal
$\widetilde{n}_i$ and $\widetilde{n}_j$ directions, as well as
on-shell conditions for $p_i$ and $p_j$.
They are given in Appendix~\ref{app:pl-mapping}.

\subsubsection{Emission probability}
As explained above, to correctly predict mass-related NLL corrections, showers are required to reproduce the correct matrix element in both the soft and quasi-collinear limits.
Let us first discuss the soft limit of the shower emission probability. Using the map detailed in Eq.~\eqref{eq:pl-map} to evaluate the matrix element~\eqref{eq:dP-soft-limit} and phase space~\eqref{eq:antennaphsp} for the emission of a soft gluon, we arrive at 
\begin{equation} \label{eq:pl-soft}
 \dd\mathcal{P}_\text{soft}^{\text{PL}} =
  \frac{4\widetilde\kappa^2_{ij}}{(1+\widetilde\kappa_{ij} - \widetilde\mu_i - \widetilde\mu_j)^2} \frac{1}{D_i^2}\frac{1}{D_j^2} \dd\Phi_\textsc{PS}\,,
\end{equation}
with
\begin{align}
	\label{eq:soft-pl}
  \dd\Phi_\textsc{PS}
   = \frac{\alpha_s C_A}{2\pi}
    \frac{\dd \alpha_k}{\alpha_k}\frac{\dd \beta_k}{\beta_k}\frac{\dd\phi}{2\pi}\,, \qquad
  D_i  = 1-\frac{x_i}{1+x_j}\frac{\alpha_k}{\beta_k}\,, \qquad   D_j = 1-\frac{x_j}{1+x_i}\frac{\beta_k}{\alpha_k}\,.
\end{align}
Note that to derive this expression, we have made use of the large-$N_c$ limit, and imposed that the emission is soft, but have not made any assumptions on the value of the emitter/spectator masses. In the massless case, $D_{i,j}\to 1$ and we recover the familiar eikonal emission probability appropriate for the emission of soft gluon.

For finite masses, the factors $D_i^{-2}$ and $D_j^{-2}$ suppress emissions in the quasi-collinear region, thereby reproducing the expected dead-cone effect. 
In particular, with $\alpha_k/\beta_k = (\tilde s_j/\tilde s_i)e^{2\bar\eta}$ (see Eq.~\eqref{eq:akbkdefn}), the suppression depends explicitly on the emission rapidity $\bar\eta$, as expected. 
Furthermore, in the quasi-collinear limit, we see that the prefactor reduces to $\tilde\mu_i$.

In the quasi-collinear limit, the light-cone momentum fraction along the spectator direction and the squared mass of the emitter and emitted particles are considered small. In this regime, the emission rate for a gluon emission off a heavy-quark becomes
\begin{align}
  \label{eq:pl-collinear}
  \dd\mathcal{P}_{\text{coll},Q \to Qg}^{\text{PL}}
  & =
    \frac{(1-\mu_j)^2\beta_k}{(1-\mu_j)^2\beta_k+\mu_i\alpha_k}
    \left[
    \frac{(1-\mu_j)^2(1-\alpha_k)\beta_k}{(1-\mu_j)^2\beta_k+\mu_i\alpha_k} + \frac{\alpha_k^2}{2}
    \right] \dd\Phi_\textsc{PS}\, .
\end{align}  
We define the longitudinal momentum fraction as
\begin{equation}
  \label{eq:zi-def}
  z_i
  = \frac{p_k\cdot \widetilde{n}_j}{\widetilde p_i \cdot \widetilde{n}_j}
  = \frac{1+x_i+x_j}{1+x_j}\alpha_k\,.
\end{equation}
Parametrically, $x_i \ll 1$ in the quasi-collinear limit, so one could also directly use $z_i = \alpha_k$. 
However, the expression in Eq.~\eqref{eq:zi-def} is also correct in the limit when the pre-branching emitter mass $\widetilde{m}_i$ is of the same order of the dipole mass $\widetilde{m}_{ij}$.
Substituting Eq.~\eqref{eq:zi-def} into Eq.~\eqref{eq:pl-collinear} and including the soft limit we obtain the following emission rate for $Q\to Qg$ splittings, with $j$ a spectator particle 
\begin{align}
  \label{eq:pl-acceptances-QQg}
  \dd\mathcal{P}_{Q(j) \to Qg(j)}^{\text{PL}}
  & =
    \frac{4\widetilde\kappa_{ij}^2}{(1+\widetilde\kappa_{ij}-\widetilde\mu_i-\widetilde\mu_j)^2}{\color{lightblue}}
    \frac{1}{D_iD_j^2}
    \left[\frac{1-z_i}{D_i}+\frac{z_i^2}{2}\right]
    \dd \Phi_\textsc{PS}\,.
\end{align}
Following the same procedure, we can also derive the emission rate for $g\to Q\bar{Q}$ splittings,
\begin{align}
\label{eq:pl-acceptances-gQQ}
  \dd\mathcal{P}_{g(j)\to Q\bar Q(j)}^{\text{PL}}
  & = \frac{T_R}{C_A}
  z_i
  \left[(1-z_i)\left(1-\frac{k_t^2z_i}{k_t^2+m_i^2}\right) - \frac{w_{q\bar{q}}}{2}(1-2z_i) \right]\dd\Phi_\textsc{ps}\,.
\end{align}
The factor $w_{q\bar q}$ in the above expression reflect the freedom in partitioning the full $g\to Q\bar{Q}$ splitting probability over the two dipoles that contribute to the gluon.
It can take any value between 0 and 1, with 0 our default as in the massless case.
An important sanity check of the previous expressions is that they reproduce the correct soft limit, namely Eq.~\eqref{eq:soft-pl}, when $z_i\to 0$. 

One last comment refers to the modification of the splitting probability for gluon emissions off $g-Q$ dipoles, i.e.\ when the spectator is massive. 
In this case, the $g\to gg$ splitting probability needs to be modified, so as to include the correct eikonal emission probability for a heavy quark to radiate a gluon.
The emission rate for such dipoles becomes
\begin{align}
  \label{eq:pl-acceptances-ggg}
  \dd\mathcal{P}_{g(j)\to gg(j)}^{\rm PL}
  & =
    \frac{1}{D_j^2} \frac{1+(1-z_i)^3 + w_{gg}z_i(1-2 z_i)}{2}\dd\Phi_\textsc{PS}\,,
\end{align}
where $w_{gg}$ again reflects the freedom to distribute the $g\to gg$
splitting function over the two dipoles that contribute to the gluon
that branches.

\subsection{\pg}
\label{sec:pg-def}
\subsubsection{Kinematic map}
The \pg shower is formulated with a global recoil scheme, where the transverse momentum generated in a branching is shared among all particles in the event.
The longitudinal momentum imbalance is absorbed in a dipole-local way.
A detailed derivation of the mapping coefficients can be found in Appendix~\ref{app:pg-mapping}, whereas here we will outline only the main steps.

The first step consists of introducing an intermediate set of post-branching momenta, $\bar{p}_l$, in such a way that only the invariant mass of the whole event is conserved.
The intermediate set of post-branching momenta, $\bar{p}_l$, are defined as
\begin{subequations}
\label{eq:pgmap}
\begin{align}
    \bar{p}_k^\mu & = a_k \widetilde n_i^\mu + b_k \widetilde n_j^\mu-  k_t n_\perp^\mu\,,\\
  \bar{p}_i^\mu & = a_i \widetilde n_i^\mu + b_i \widetilde n_j^\mu\,,\\
  \bar{p}_j^\mu & = a_j \widetilde n_i^\mu + b_j \widetilde n_j^\mu\,,\\
  \bar{p}_l^\mu & = \widetilde p_l^\mu \quad \text{for } l\neq i,j,k\,,
\end{align}
\end{subequations}
with $n_\perp^\mu$ defined in Eq.~\eqref{eq:nperp}, and $k_t$ given in Eq.~\eqref{eq:kt2fornperp}.
In the massless case~\cite{Dasgupta:2020fwr,FerrarioRavasio:2023kyg}, we define the non-zero mapping coefficients as
\begin{equation}
	\label{eq:massless-pg}
a_k = (1+\delta)\alpha_k, \quad
 b_k = (1+\delta)\beta_k, \quad 
a_i = (1+\delta)(1-\alpha_k),\quad
 b_j = (1+\delta)(1-\beta_k), \quad k_t = (1+\delta)\kappa_t,
\end{equation}
with $\kappa_t^2 = \alpha_k \beta_k \widetilde{s}_{ij}$ being the squared shower effective transverse momentum introduced in Eq.~\eqref{eq:kappatdefn}.
An important difference with respect to the \pl mapping is that the Sudakov coefficients $a_k$ and $b_k$ of the emitted particle $k$ do not coincide with $\alpha_k$ and $\beta_k$, which are directly related to the shower variables $v$ and $\bar{\eta}$ (see Eq.~\eqref{eq:akbkdefn}).
Indeed, all mapping coefficients contain an overall factor $\rl = 1+\delta$, 
to ensure that the total centre-of-mass energy is preserved.

When handling massive partons, a naive rescaling of $\rl=(1+\delta)$
of all coefficients, like done above in Eq.~\eqref{eq:massless-pg},
will generally spoil the on-shell condition for massive particles. 
To amend this we introduce a second rescaling factor $r_k$, and change the relationship between $\kappa_t$ and $k_t$. 
This modification has the property that it reduces to the same massless \pg shower of Refs.~\cite{Dasgupta:2020fwr,FerrarioRavasio:2023kyg} when all particles are massless.
More concretely, we define the $\bar p^\mu_k$ coefficients as 
\begin{align}
  \label{eq:akbkPG}
  &a_k = \left(1+\widetilde{\kappa}_{ij}\delta \right) r_k \alpha_k, 
\quad b_k = \left(1+\widetilde{\kappa}_{ij}\delta \right) r_k \beta_k,\quad
k_t = (1+\widetilde{\kappa}_{ij}\delta)\sqrt{\kappa_t^2-m_k^2}\,.
\end{align}
One can determine $r_k$ as a function of $\alpha_k$, $\beta_k$ and $\delta$ by imposing $p_k^2=m_k^2$ (see Eq.~\eqref{eq:pg-rk}). 
\pg conserves the longitudinal momentum in a dipole-local way. This means that
we require that the sum of the longitudinal components satisfies a simple relation, namely
\begin{equation}
  a_i+a_j+a_k = b_i+b_j+b_k=\frac{1+\widetilde{\kappa}_{ij}\delta}{\widetilde{\kappa}_{ij}},
\end{equation}
where the factor $\frac{1}{\widetilde{\kappa}_{ij}}$ is defined in
Eq.~\eqref{eq:kallen}.
With these relations, one can find the factor $\delta$ that ensures the total centre-of-mass energy is preserved (see Eq.~\eqref{eq:pg-rescaling}).
Requiring that $a_i b_i \widetilde{s}_{ij} = m_i^2$ and $a_j b_j \widetilde{s}_{ij} = m_j^2$ enables us to determine all the coefficients as a function of $\alpha_k$, $\beta_k$ (see Eq.~\eqref{eq:pg-mapping}).

Finally, to restore the total four-momentum of the event we apply a Lorentz boost to all particles, i.e.
\begin{equation}
	p_l^\mu = \Lambda^{\mu}_{\phantom{\mu}\nu} \bar{p}_l^\mu\,,
\end{equation}
with
\begin{equation}
	\Lambda^{\mu}_{\phantom{\mu}\nu} = g^{\mu}_{\phantom{\mu}\nu} + 2 \frac{\widetilde Q^{\mu} \bar{Q}_\nu}{\widetilde Q^2}-2 \frac{(\widetilde Q+\bar{Q})^{\mu} (\widetilde Q+\bar{Q})_\nu}{(\widetilde Q+\bar{Q})^2}\,,
\end{equation}
where $\widetilde{Q}^\mu = \sum_l \tilde p_l^\mu$ and $\bar{Q}^\mu = \sum_l \bar p_l^{\mu}$.

We finalise our presentation on the \pg kinematic map with a discussion
of some properties of the rescaling factors. First, $r_k$ is different from 1 only if $k$ is a massive parton, i.e.\ for $g\to Q\bar{Q}$ splittings. Secondly, in the limit of a soft or collinear emission, we have $\rl \to 0$ and $r_k \to 1$. Note that with the definition in Eq.~\eqref{eq:akbkPG}, we have $a_k/b_k=\alpha_k/\beta_k$, as in the original massless formulation, so that the rapidity of the emitted particle in the event frame is directly related to the shower variable $\bar{\eta}$, see Eq.~\eqref{eq:akbkdefn}.
Furthermore, the absolute value of the transverse momentum of the emitted particle has a simple relation to the shower effective transverse momentum $\kappa_t^2$.

\subsubsection{Emission probability}
Using the \pg map to evaluate the matrix element and phase space
  for the emission of a soft gluon, we find the same expression as for
  \pl, given in Eqs.~\eqref{eq:pl-soft}
  and~\eqref{eq:soft-pl}.
However, in the quasi-collinear limit, the emission probability becomes
\begin{equation}
	\dd \mathcal{P}_{{\rm col}, Q\to Qg}^{\rm PG} =
	\frac{1}{\bar{D}_i}\left(\frac{1-\alpha_k}{\bar{D}_i} + \frac{\alpha_k^2}{2}\right)\dd\Phi_{\rm PS}\,,
\end{equation}
with $\dd\Phi_{\rm PS}$ the phase-space measure defined in
Eq.~\eqref{eq:soft-pl} and
\begin{equation}
  \bar{D}_i
  = 1 - \frac{x_i}{1 + x_j} \frac{\alpha_k}{\beta_k(1-\alpha_k)^2}
  = 1 + \frac{1 + {\mu}_i - {\mu}_j - \widetilde{\kappa}_{ij}}{1 +
    {\mu}_i - {\mu}_j + \widetilde{\kappa}_{ij}}
  \frac{\alpha_k}{\beta_k(1-\alpha_k)^2}
  .
\end{equation}
Combining the soft and collinear limits, we write the full emission probability for a $Q\to Qg$ splitting as 
\begin{align}
 	\dd \mathcal{P}_{Q\to Qg}^{\rm PG} &=
 	\frac{4\widetilde{\kappa}_{ij}^2}{(1 + \widetilde{\kappa}_{ij} - {\mu}_i - {\mu}_j)^2} \frac{1}{\bar{D}_i}\frac{1}{\bar{D}_j^2} \left(\frac{1-z_i}{\bar{D}_i} + \frac{z_i^2}{2}\right)  \dd\Phi_{\rm PS} \,,
\end{align}
with $z_i$ as defined in Eq.~\eqref{eq:zi-def}, and
\begin{equation}
  \bar{D}_j
  =1 - \frac{x_j}{1 + x_i} \frac{\beta_k}{\alpha_k(1-\beta_k)^2}
  =1 + \frac{1 + {\mu}_j - {\mu}_i - \widetilde{\kappa}_{ij}}{1 + {\mu}_j - {\mu}_i + \widetilde{\kappa}_{ij}} \frac{\beta_k}{\alpha_k(1-\beta_k)^2}.
\end{equation}
For massless legs, $\bar{D}_i$ and $\bar{D}_j$ reduce to 1 and we recover the familiar massless splitting function for $q\to qg$ splittings.

As discussed for \pl, the $g\to gg$ emission rate has to be modified
to accommodate massive spectators. For \pg we include the dead-cone
suppression when the spectator is a heavy quark by including the factor
$1/\bar{D}_j^2$ in the splitting probability, so that the emission
rate becomes
\begin{align}
		\label{eq:pg-acceptances-gg-gg}
		\dd\mathcal{P}_{g\to gg}^{\rm PG}
		= 
		\frac{1}{\bar D_j^2} \frac{1+(1-z_i)^3 + w_{gg}z_i(1-2 z_i)}{2} \dd\Phi_\textsc{PS}\,.
\end{align}
Lastly, for $g\to Q\bar{Q}$ we use
\begin{align}
\dd\mathcal{P}_{g\to Q\bar{Q}}^{\rm PG} &= f_{\rm split}\frac{T_R}{C_A} \frac{(1-\alpha_k)^2 [(1-\mu_j)^2 \kappa_t^2 + \mu_i]}{\mu_i + (1-\alpha_k)^2(1-\mu_j)^2 \kappa_t^2} \nonumber \\ &\times \alpha_k \left[\alpha_k^2 + (1-\alpha_k)^2 + \frac{2\mu_i \alpha_k(1-\alpha_k)}{\mu_i + (1-\alpha_k)^2(1-\mu_j)^2 \kappa_t^2}\right]\dd\Phi_{\rm PS} \,,
\label{eq:PgQQPGSym}
\end{align} 
where the factor
\begin{equation}
f_{\rm split} = \frac{(1-\alpha_k)^2-w_{q\bar{q}}(0.5-\alpha_k)}{(1-\alpha_k)^2+\alpha_k^2}\,,
\end{equation}
takes care of the symmetry factor as the gluon that splits belongs to two dipoles. 
This, together with the factor $w_{q\bar q}$ is chosen such that the massive splitting kernel smoothly reduces to the massless version, and our default value is $w_{q\bar{q}} = 0$. 

\subsection{Flavour thresholds in the effective coupling}
\label{sec:kcmw-threshold}
In Eq.~\eqref{eq:singleleg-prob} we have implied that the shower emission probability depends on the effective coupling $\alpha_s^{\rm eff}$. 
At NLL this effective coupling is defined as
\begin{align}
\alpha_s^{{\rm eff}, (n_{\!f})}(\kappa_t) = \alpha^{(n_{\!f})}_s(\kappa_t)\left(1 + \frac{\alpha^{(n_{\!f})}_s(\kappa_t)}{2\pi}K^{(n_{\!f})}_{\rm CMW}\right)\,,
\end{align} 
with $\alpha^{(n_{\!f})}_s(\kappa_t)$ the two-loop $\overline{\rm MS}$
coupling and $n_f$ the number of light flavours. The running of
$\alpha^{(n_{\!f})}_s(\kappa_t)$ is controlled by the $\beta$ function,
with coefficients given by
\begin{subequations}
\begin{align}
  \label{eq:beta0-def}
  b_0^{(n_{\!f})} & = \frac{11}{12\pi}C_A - \frac{1}{3\pi} T_R n_f\,,\\
  b_1^{(n_{\!f})} & = \frac{17}{24\pi^2}C_A^2 - \frac{5}{12\pi^2}C_AT_Rn_f-\frac{1}{4\pi^2}C_FT_Rn_f\,.
\end{align}
\end{subequations}
The soft cusp anomalous dimension $K^{(n_{\!f})}_{\rm CMW}$ also depends on the number of active flavours, and is given by
\begin{align}
K^{(n_{\!f})}_{\rm CMW} = \left(\frac{67}{18}-\frac{\pi^2}{6}\right)C_A - \frac{10}{9}T_R n_f\,.
\end{align}
When crossing a heavy-quark threshold, $n_{\!f}$ changes by one
unit. For the $\overline{\rm MS}$ coupling, we adopt the customary
  prescription whereby the coupling is continuous at the heavy quark
  mass ($\kappa_t = m_Q$), i.e.\ $\alpha_s^{(n_{\!f})}(m_Q)=\alpha_s^{(n_{\!f}-1)}(m_Q)$. 
The flavour thresholds are then taken at a scale $\kappa_t =
\mu^{(n_{\!f})}$ implicitly defined through 
\begin{align}
  \label{eq:as-th}
	\alpha_s^{{\rm eff}, (n_{\!f})}(\mu^{(n_{\!f})}) = \alpha_s^{{\rm eff},(n_{\!f}-1)}(\mu^{(n_{\!f})})\,.
\end{align}
While Eq.~\eqref{eq:as-th} may be solved numerically to find $\mu^{(n_{\!f})}$, it is instructive to look at its solution when expanding the running of $\alpha_s^{{\rm eff},(n_{\!f})}$.
This allows us to understand the difference in threshold values when $K^{(n_{\!f})}_{\rm CMW}$ is turned on. 
Up to $\mathcal{O}(\alpha_s)$, the effective shower coupling agrees with the {\rm MS} one, so the requirement that $\alpha_s^{{\rm eff},(n_{\!f})}(\mu^{(n_{\!f})}) = \alpha_s^{{\rm eff},(n_{\!f}-1)}(\mu^{(n_{\!f})})$ gives $\mu^{(n_{\!f})} = m_Q$. 
Expanding the effective coupling to $\mathcal{O}(\alpha_s^2)$ we find that the continuity requirement translates into a shift of the threshold scale according to
\begin{align}
  \label{eq:threshold-shift}
\ln m_Q \to \ln \mu^{(n_{\!f})} = \ln m_Q +  \frac{1}{4\pi} \frac{K_{\rm CMW}^{(n_{\!f})} - K_{\rm CMW}^{(n_{\!f}-1)}}{b_0^{(n_{\!f})} - b_0^{(n_{\!f}-1)}} = \ln m_Q  + \frac{5}{6}\,.
\end{align}
Since the factor in the second term of this equation is positive we see that the effective coupling switches to use a smaller number of light flavours at a higher scale than the $\overline{\rm MS}$ coupling would. 
Note that, strictly speaking, there is no physical requirement for the effective coupling to be continuous across flavour thresholds. 
For instance, an alternative prescription would be to enforce continuity of $\alpha_s^{n_{\!f}}(\kappa_t)$ at the physical mass scale $\kappa_t = m_Q$ and then include a compensating finite matching correction that coincides with the second term in Eq.~\eqref{eq:threshold-shift}. 
These prescriptions differ only by terms beyond our targeted NLL accuracy.

\subsection{Preserving the logarithmic behaviour of the massless showers}
\label{sec:log-acc-massless}
In addition to enhancing the logarithmic accuracy of the showers whenever quark masses are relevant, we also aim to make sure that the original massless accuracy of our showers is preserved. 
In particular, for \pg we retain NNLL accuracy for global observables~\cite{vanBeekveld:2024wws}, while for \pl\ we preserve next-to-next-to-double logarithmic (NNDL) accuracy~\cite{Hamilton:2023dwb}. 
To do so, we include a series of ingredients, which are explained in this section.

First, to achieve NNDL accuracy for massless quarks we implement next-to-leading order multiplicative matching using the massless matrix element as described in Ref.~\cite{Hamilton:2023dwb}. 
More specifically, multiplicative matching in the massless case modifies the differential cross section $\dd\sigma$ as
\begin{align}\label{eq:mult-matching}
\dd\sigma_{\rm mult} = \bar{B}(\Phi_B) \left[S^{m=0}_{\rm PS}(v) \times \frac{R_{\rm PS}^{m=0}(\Phi)}{B_0(\Phi_B)} \times \frac{R(\Phi)}{R_{\rm PS}^{m=0}(\Phi)} \dd\Phi \right] \times I_{\rm PS}(v,\Phi)\,.
\end{align}
In the previous expression, $\Phi$ is the full Born- plus-one phase space, $\Phi_B$ the Born phase space, $R$ the real matrix element, $B_0$ the Born matrix element, and $\bar{B}$ the NLO normalisation factor.
The massless shower Sudakov is denoted by $S^{m = 0}_{\rm PS}$ and reads
\begin{align}\label{eq:massless-sudakov-matching}
S_{\rm PS}^{m = 0}(v) = \exp \left[-\int_v \frac{R_{\rm PS}^{m=0}(\Phi)}{B_0(\Phi_B)} \dd\Phi_{\rm rad}\right],
\end{align}
with the radiation phase-space defined through $\dd\Phi = \dd\Phi_B  \dd\Phi_{\rm rad}$, and $v$ the value for the shower ordering variable. 
The factor $I_{\rm PS}$ represents the iteration of the shower after the first emission has been accepted.
The matrix element $R_{\rm PS}^{m=0}$ represents the shower matrix
element, which is constructed such that $R_{\rm PS}^{m=0}$ shares the
same soft or collinear singularities as $R$.
To preserve the behaviour of the massive shower in the dead-cone region, we perform the following replacement
\begin{equation}
  \frac{R_{\rm PS}^{m=0}(\Phi)}{B_0(\Phi_B)}
  \to
  \frac{R_{\rm PS}(\Phi)}{B_0(\Phi_B)} 
\end{equation}
both in~\eqref{eq:mult-matching} and in \eqref{eq:massless-sudakov-matching}.
In addition, we evaluate the matrix element with \emph{demassified} emission kinematics, 
where we preserve the direction and the energy of the massive particles, but change the magnitude of the three momentum to make them massless. 
We show a check of this procedure in Sec.~\ref{sec:me-tests}.

The second ingredient that needs to be adjusted are the NNLL Sudakov effects, which are currently only available for the \pg shower. 
To this end, we evaluate the various NNLL corrections in using a variable $n_f$. 
This affects the $\beta$-coefficients of the running-coupling, the cusp anomalous dimension, the hard-collinear constants, and other NNLL ingredients entering the Sudakov exponent (i.e., \ the drifts).
The value of the threshold scale remains unmodified with respect to the NLL case, meaning we use the threshold scale of Eq.~\eqref{eq:threshold-shift}.

Regarding the double-soft matrix-element corrections of Ref.~\cite{FerrarioRavasio:2023kyg}, which like the NNLL corrections are currently only implemented for \pg, we ensure that the standard double-soft result is recovered when masses are negligible, while avoiding spurious distortions in regions dominated by mass effects. 
In the massless case, double-soft corrections are implemented by introducing an additional double-soft acceptance probability to Eq.~\eqref{eq:shower-prob} for the second emission onwards that reads
\begin{align}
p_{\rm accept}^{\rm DS} = \frac{|M_{\rm DS}|^2}{\sum_h |M^{m=0}_{{\rm PS},h}|^2}\,,
\end{align}
where $|M_{\rm DS}|^2$ is the double-soft matrix element and $\sum_h |M_{{\rm PS},h}|^2$ runs over all the effective squared shower matrix elements, evaluated in the massless double-soft limit, that have contributed to the specific kinematic configuration under consideration (see Ref.~\cite{FerrarioRavasio:2023kyg} for more details).
Correctly implementing the double-soft reweighting factor with masses, which lies beyond the scope of this work, would require an implementation of the massive double-soft matrix element, in addition to a correction of the shower effective double-soft matrix elements. 
Instead, here we require that the double-soft acceptance goes to one if the particles are well separated in the Lund plane, and that it reduces to the massless double-soft acceptance in a region where masses are parametrically irrelevant. 
This is done by evaluating all matrix elements in the variable number-of-flavours scheme, where the threshold scale is set by Eq.~\eqref{eq:threshold-shift}.
The effective double-soft squared matrix elements are evaluated using massless matrix elements to ensure that they map onto the double-soft matrix element in strongly-ordered limits, but we keep massive momenta for the construction of the invariants that enter these matrix elements.  
Moreover, double-soft corrections in channels involving $g\to Q\bar{Q}$ splittings are vetoed when the transverse momentum is below the threshold scale, i.e. $\kappa_t < m_Q$.
We checked that the shower behaves the same when setting $n_f = 4$ directly versus working in the variable-flavour scheme, but with one quark mass being larger than the centre-of-mass energy, while the others all massless. 

\section{Fixed-order validation}
\label{sec:fo-val}
\subsection{Phase-space contours for two emissions}
\label{sec:contours}
A necessary condition for a shower to be NLL accurate is to reproduce
the correct squared matrix element for multiple emissions when those emissions are well-separated. 
However, the implementation of the correct emission probability is not
enough to ensure a correct radiation pattern. A second requirement is
that an emission that is well separated in phase space from other prior emissions, should not alter the kinematics of those prior emissions~\cite{Dasgupta:2018nvj}. 

To understand whether this requirement is fulfilled, we follow the recipe introduced in Ref.~\cite{vanBeekveld:2022zhl}.
For these tests, we do not include matching or any of the higher-order
corrections discussed in Sec.~\ref{sec:log-acc-massless} (which
would anyway only affect the emission's weight).
We first generate a single emission off a Born configuration with a fixed momentum $\widetilde{k}_1^\mu$. Then we add a second emission with a fixed value of the evolution variable, and check the new value of the momentum of the first emission, $k_1^\mu$. If $k_1^\mu$ and $\widetilde{k}_1^\mu$ differ significantly in a logarithmically-enhanced region of phase space for $k_2^\mu$, this will lead to a failure of the shower to be NLL accurate. In what follows, we test this condition for the \panscales showers discussed in Secs.~\ref{sec:pl-def} and \ref{sec:pg-def}, and for our implementation of the \dire shower~\cite{Hoche:2015sya}, so as to have a comparison to a shower that does not fulfil this requirement.

\begin{figure}[t!]
	\centering
	\includegraphics[width=0.48\textwidth,page=1]{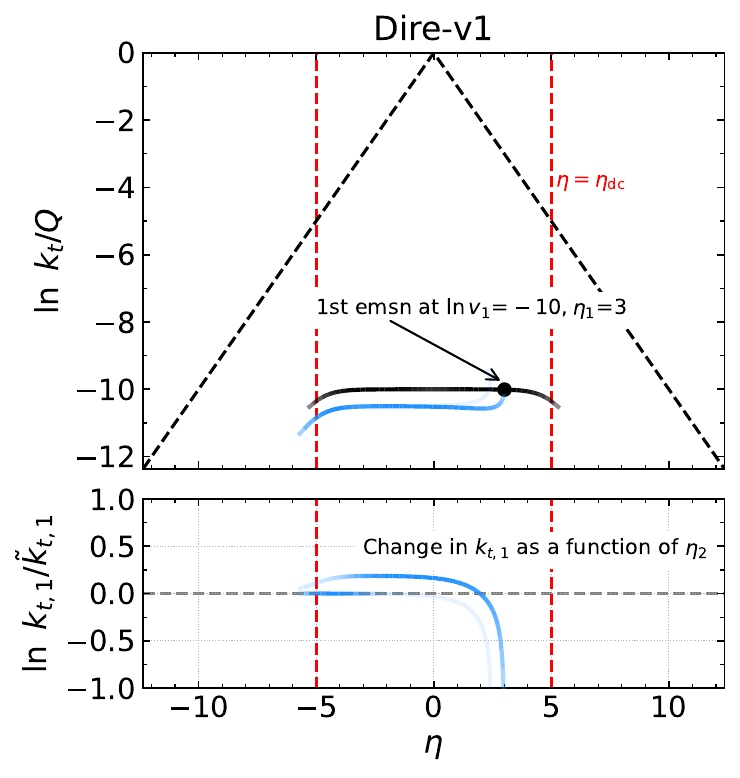}
	\includegraphics[width=0.48\textwidth,page=2]{figures/plot-contours-pg00.pdf}
	\includegraphics[width=0.48\textwidth,page=2]{figures/plot-contours-pg05.pdf}
	\includegraphics[width=0.48\textwidth,page=1]{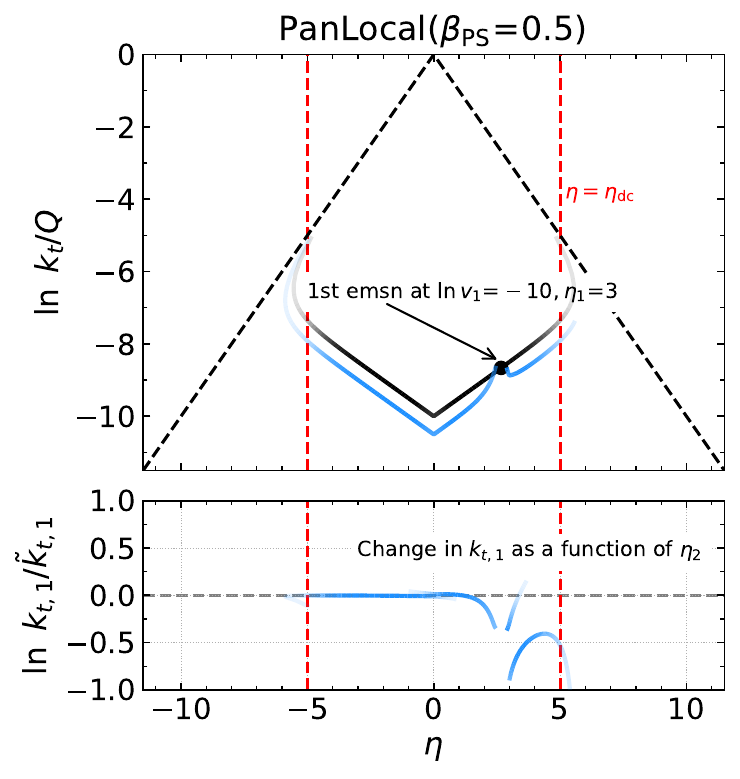}
	\caption{Phase-space contours for \dire (top, left), \pg with $\betaps=0$ (top, right) and $\betaps=0$ (bottom, left), and \pl with $\betaps=1/2$ (bottom, right). The contours for the first and second emissions are shown in black and blue, respectively, and the opacity of the line is indicative of the shower weight with which that emission would be produced. The first emission is generated with $\ln v_1/Q = -10$ and with $\eta_1 \approx 3$, and the second emission with $\ln v_2/Q = -10.5$. The red, vertical dashed lines at $\eta=5$ represent the dead-cone rapidity. In the bottom panel we show the logarithm of the ratio between the transverse momentum of the first emission after ($k_{t,1}$) and before the second emission took place ($\tilde k_{t,1}$), which is expected to be zero (dashed line) except when the two emissions are close in rapidity.}
	\label{fig:emission-contours-vanilla}
\end{figure}

\begin{figure}[t]
	\centering
	\includegraphics[width=0.48\textwidth,page=1]{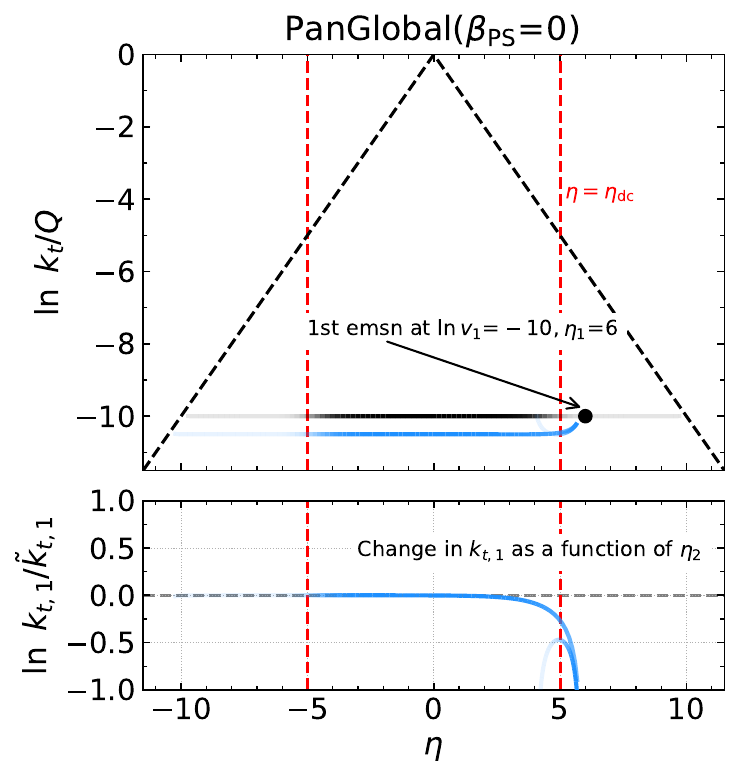}
	\includegraphics[width=0.48\textwidth,page=1]{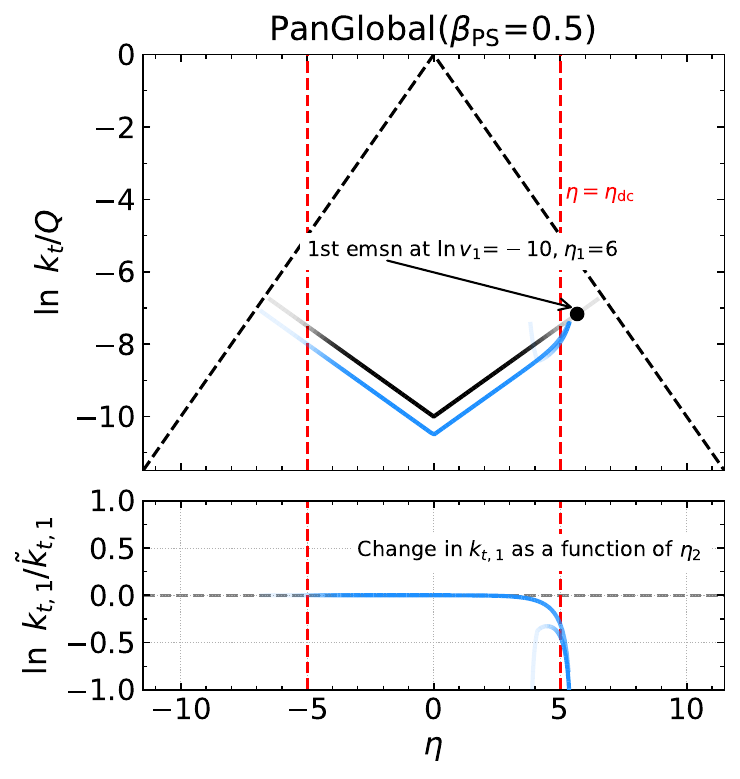}
	\includegraphics[width=0.48\textwidth,page=2]{figures/plot-contours-pl05.pdf}
	\caption{Same as Fig.~\ref{fig:emission-contours-vanilla} but with the first emission being in the dead-cone region, $\eta_1 \approx 6$.}
	\label{fig:emission-contours-higher-mass}
\end{figure}

The Born configuration consists of a $Q\bar{Q}$ pair produced in $e^+e^-$ annihilation at a centre-of-mass energy $Q$, with the heavy quark mass denoted $m_Q$.
We then generate an emission at a given shower variable $\ln v_1$ and
azimuthal angle $\phi_1$, varying the auxiliary $\bar \eta_1$
shower variable.
This results in the black contours in
Figs.~\ref{fig:emission-contours-vanilla}
and~\ref{fig:emission-contours-higher-mass}.
The opacity of any given line represents the shower weight for that kinematic point.
We then fix a specific value of $\bar\eta_1$ for the first emission (represented by the black dot), and generate a second emission at $\ln v_2<\ln v_1$, whose contour is instead coloured in blue.
The contours themselves are defined in terms of the Lund~\cite{Dreyer:2018nbf, Andersson:1988gp} variables $(\ln k_t, \eta)$.
To construct these variables, we cluster the event into two jets using the Cambridge jet-algorithm~\cite{Dokshitzer:1997in} using the winner-takes-all recombination scheme, where the modulus of the three momentum is used to determine the `winner'~\cite{Larkoski:2014uqa}. 
If only one emission is present, we take the last clustering step and compute 
\begin{equation}
  \label{eq:lund-variables}
  k_t = \min(E_i, E_j)\,\sin \theta_{ij}\,,
  \qquad
  \eta = -\ln \tan \frac{\theta_{ij}}{2}\,.
\end{equation}
where $\theta_{ij}$ is the angle between the two subjets and $E_{i,j}$ are their corresponding energies. 
For the two-emission case, we only consider primary emissions, which means that both of the softer emissions must have clustered with one of the Born quarks instead of clustering together in one jet.
Then, we undo the last two clustering steps, record $k_t$ and $\eta$ as defined above for each declustering, and plot ($\ln k_t,\eta$) for the declustering associated with the second emission.

In Fig.~\ref{fig:emission-contours-vanilla}, we show results where the
first emission happens at $\ln v_1/Q = -10$, and the mass of the quark
is chosen such that $\eta_{\rm dc} = 5$, with $\eta_\text{dc}=\ln\frac{Q}{m_Q}$.
We then fix a kinematic configuration for the first emission such that $\eta_1 \simeq 3$~(black dot) and we scan over the second emission rapidity, fixing $\ln v_2/Q = -10.5$.

We will start with a LL shower to clearly illustrate the kind
    of features that appear in the presence of NLL failures. 
We will use our custom implementation of the \dire shower (top of  Fig.~\ref{fig:emission-contours-vanilla}, left panel) to illustrate this. 
Focusing on the black line first, i.e.\ the first-emission contour, we see that this line quickly disappears for rapidities around and beyond the dead-cone boundary.
While emissions inside the dead-cone region are power-suppressed, they
are not forbidden.
Thus, we observe that the \dire shower does not fully populate the phase space. In addition, the transverse momentum of the first emission is affected by the second emission even when the pair of emissions is widely separated in rapidity. This feature, whose origin is the same as the corresponding feature found in the massless case~\cite{Dasgupta:2018emf}, is the source of NLL accuracy breaking. 

The \panscales showers do generate emissions inside the dead-cone region with their probability naturally suppressed compared to that in the bulk of the Lund plane. 
The contours for the second emission display a number of properties in the deep collinear region ($\eta_2>\eta_{\rm dc}$), none of which affect logarithmic accuracy, but are worth to point out nonetheless. 
First of all we see that once a first gluon has been emitted off a
certain leg, the phase space for a second gluon off that same leg does
not extend all the way to the dashed black line. 
This is due to the fact that an emission collinear to a heavy leg reduces the phase space available for subsequent collinear emission, as the heavy leg needs to have energy at least equal to its mass.
A second observation is that also in this deep collinear region, the transverse momentum of the first gluon is significantly altered, even when the two emissions have a large rapidity separation. For instance, in the \pg($\betaps=0.5$) case we observe that $\ln k_{t,1}$ and $\ln \tilde k_{t,1} $ differ significantly for $\eta_2>\eta_{\rm{dc}}$ and $|\eta_2-\eta_1|>2$. 
This only occurs when the second emission is radiated deep inside the dead-cone and is due to both \pl and \pg showers conserving longitudinal momentum in a dipole-local way. Nevertheless, we found that the shower's weight for such configurations vanishes as $1/k^4_t$ and thus this effect does not spoil NLL accuracy.   

In Fig.~\ref{fig:emission-contours-higher-mass}, we show the emission contours when the first emission occurs in the dead-cone region with $\eta_1 \approx 6$. When the second emission is radiated off the opposite leg, no new features appear. 
In turn, we observe an uprising trend of the second emission contour for $\eta_2 < \eta_1$, e.g.\ for $\eta_2 \approx 4$ in the \pl($\betaps=0.5$) case. 
This is due to the fact that, after the first emission, the quark received a substantial longitudinal recoil and now it has a rapidity smaller than the pseudorapidity of the declustering, $\eta_1$. 
Note that secondary emissions off the dipole leg corresponding to the first emission are not shown in this primary Lund plane. 

\subsection{Matrix element tests up to $\mathcal{O}(\alpha_s^2)$}
\label{sec:me-tests}
We now turn to the test of the shower's branching probability and colour assignment. 
To this end, we start off with a fixed Born configuration.
We then compare differentially the shower fixed-order weight (i.e.\ without the contribution of its Sudakov), $d\sigma_{\rm PS}$, with the exact QCD result, $d\sigma_{\rm exact}$, for a series of final state configurations including up to two emissions.
Our reference QCD result is obtained by generating the four-momenta of emissions with an in-house soft-collinear phase-space generator and evaluating the event weight using the relevant matrix elements provided by \codefont{MadGraph v3.6.3}~\cite{Alwall:2014hca}.

\begin{figure}
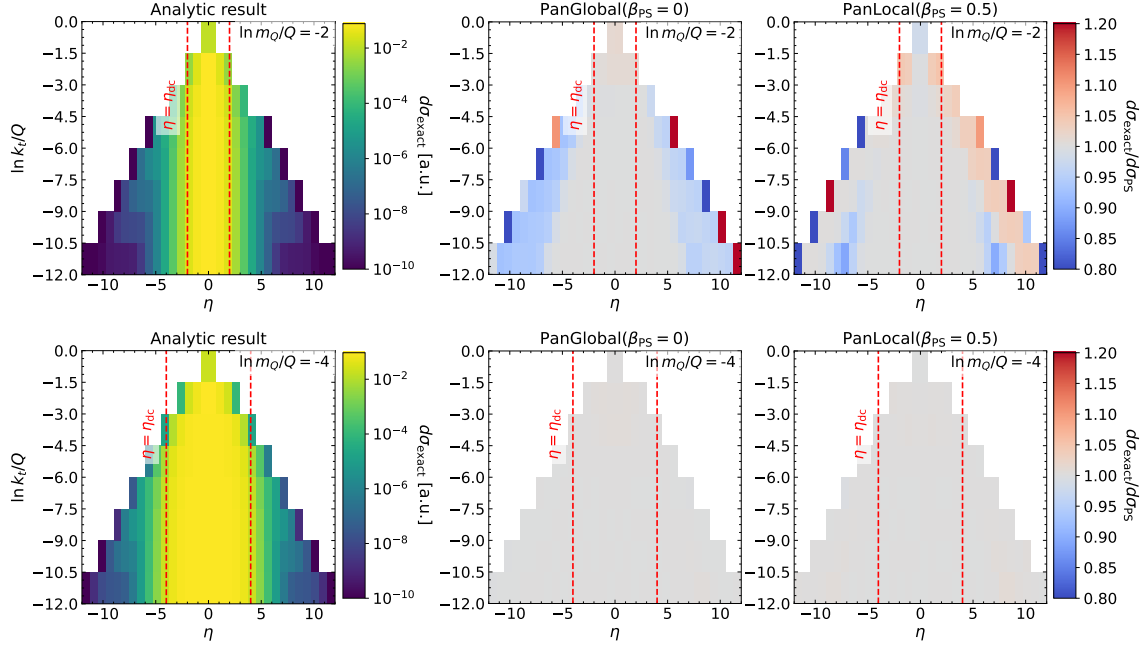

	\centering
	\includegraphics[width=\textwidth,page=5]{figures/plot_mass_checks_1emsn.pdf}
	\includegraphics[width=\textwidth,page=8]{figures/plot_mass_checks_1emsn.pdf}
	\caption{Fixed-order tests for $e^+e^-\to Q\bar{Q}g$ using two different quark masses: $\ln m_Q/Q = -2$ (top) and $\ln m_Q/Q = -4$ (bottom). The red, vertical dashed lines represent the dead-cone rapidity. First column: Lund-plane representation of the exact differential cross-section in arbitrary units. Second column: ratio between the exact differential cross-section and the shower approximation for \pg with $\betaps=0$. Third column: same as the second column but for \pl with $\betaps=0.5$. \panscales showers include matrix-element corrections in the massless limit.}
	\label{fig:one-emission-test}
\end{figure}

As for the tests described in Sec.~\ref{sec:contours}, we use the
Cambridge algorithm with winner-takes-all recombination to cluster events with one or two emissions, and we histogram the Lund coordinates associated with the last clustering step. In some cases we will consider both primary and secondary declusterings.\footnote{As before, here we follow the standard Lund prescription and classify the softer declustering as primary if the emission was clustered with one of the Born quarks, and secondary if the two emitted particles were clustered together.}

Let us begin with the one-emission tests. 
The Lund density, i.e.\ the differential cross-section binned in the Lund variables $\ln k_t$ and $\eta$, for
the process $e^+e^- \to Q \bar Q g$ is shown in the first column of Fig.~\ref{fig:one-emission-test} for two different quark masses $m_Q$ corresponding to $\eta_{\rm dc} = 2$ (top) and $\eta_{\rm dc} = 4$ (bottom). 
We observe the expected depletion of the cross-section around $\eta = \eta_{\rm dc}$. 
The ratio $d\sigma_{\rm exact}/d\sigma_{\rm PS}$ is shown in the second and third columns of Fig.~\ref{fig:one-emission-test} for \pg and \pl, respectively, where we incorporate $\mathcal{O}(\alpha_s)$ massless matrix-element corrections according to the prescription described in Sec.~\ref{sec:log-acc-massless}.
Overall, we observe excellent agreement between the shower and the exact result, except for two general regions.
The first region is associated with large values for $k_t$, which is affected by matching corrections. 
While that region is correctly described when the mass of the heavy quark becomes small relative to $Q$, for higher mass values, mass corrections in the hard matrix element for the $Q\bar{Q}g$ configuration become important and consequently the shower does not agree with the exact result in that region (as can be seen in the top row of Fig.~\ref{fig:one-emission-test}, where $m_Q\approx 0.14\,Q$).
The second region is associated with the deep quasi-collinear region. 
Again, when the quark mass becomes relatively large compared to the centre-of-mass energy, we observe, depending on the shower, up to $10\%$ deviations in a region close to the collinear boundary, where the gluon is hard and quasi-collinear to the quark.
This is due to missing power corrections of the form $k_t^2/Q^2$ that
become relevant deep inside the dead-cone region where radiation is
highly suppressed as $\exp(-4(\eta-\eta_\text{dc}))$. This explains
both the shape of the region where deviations are observed and the
fact that this effect is suppressed (as $\exp(-2\eta_\text{dc})$) when
the quark mass is small compared to the centre-of-mass energy (e.g.\
bottom row of Fig.~\ref{fig:one-emission-test} where $m_Q\approx 0.02\, Q$).
Note that the soft region of the dead cone is well described by the shower.   

\begin{figure}
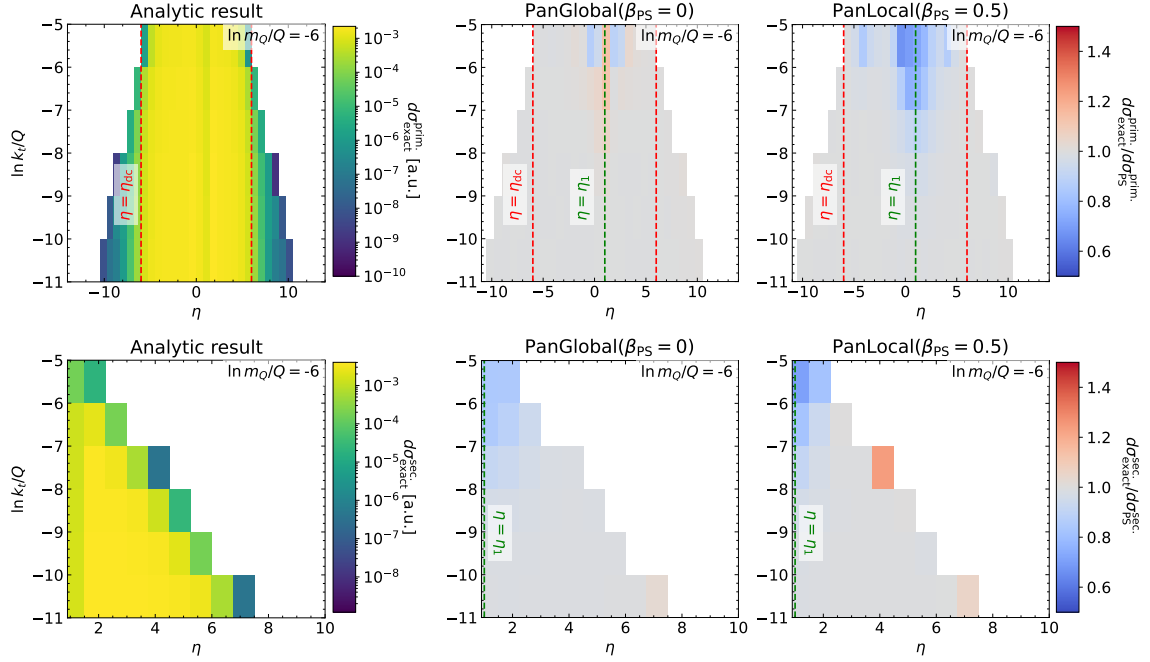

\centering
	\includegraphics[width=\textwidth,page=4]{figures/plot_mass_checks_2emsn.pdf}
 	\includegraphics[width=\textwidth,page=14]{figures/plot_mass_checks_2emsn.pdf}
\caption{Fixed-order tests for $e^+e^-\to Q\bar{Q}g_1g_2$ using $\ln m_Q/Q = -6$, as a function of the Lund coordinates $k_t$ and $\eta$ of the softer declustering.
The hardest declustering satisfies $\ln k_t \in [\ln k_{t,1} - 0.25,\; \ln k_{t,1} + 0.25\,]$ and $\eta \in [\,\eta_1 - 0.25,\; \eta_1 + 0.25\,]$,
with $\ln k_{t,1}/Q = -5$ and $\eta_1 = 1$~(dashed green line).
The red, vertical dashed lines represent the dead-cone rapidity.
The top row corresponds to the case in which both gluons cluster directly with a subjet containing either a $Q$ or a $\bar Q$, (i.e.\ is a primary clustering), while the bottom row corresponds to the case in which the first clustering is the one between $g_1$ and $g_2$ (i.e. is a secondary clustering).
$1$st column: Lund-plane representation of the exact differential cross-section in arbitrary units. $2$nd column: ratio between the exact differential cross-section and the shower approximation for \pg with $\betaps=0$. $3$rd column: same as the $2$nd column but for \pl with $\betaps=0.5$.}
\label{fig:two-emission-test-gg-1}
\end{figure}

We now examine configurations at the level of two real emissions, and
consider two processes: (a) $e^+e^- \to Q\bar{Q} g_1g_2$ and (b)
$e^+e^- \to q\bar q Q\bar{Q}$.
We sample the kinematics for two emissions, starting either from a
$Q\bar{Q}$ final state for (a), or a $q\bar{q}$ final state for (b).
In this case, we do not include the higher-order corrections described
in Sec.~\ref{sec:log-acc-massless}.\footnote{The main reason for this
  is that our corrections in Sec.~\ref{sec:log-acc-massless} are only
  valid in the large-$N_c$ limit while the tests carried out here are
  done at full $N_c$.}
We then proceed by clustering the event with Cambridge algorithm, and apply the Lund declustering procedure.
We impose that the hardest declustering satisfies $|\ln {k_t}/{k_{t,1}}|<0.25$ and $|\eta-\eta_{1}|<0.25$, where
${k_{t,1}}$ and $\eta_{1}$ are fixed reference values, and we plot the Lund variables associated with the softer declustering.
We then compute the ratio between the shower weight and the exact matrix element, $d\sigma_{\rm exact}/d\sigma_{\rm{PS}}$, as a function of the Lund coordinates of the second declustering ($k_t, \eta$). 
We plot this ratio following a Lund-style representation where we consider both primary, $d\sigma^{\rm prim.}_{\rm exact}/d\sigma^{\rm prim.}_{\rm{PS}}$, and secondary, $d\sigma^{\rm sec.}_{\rm exact}/d\sigma^{\rm sec.}_{\rm{PS}}$, splittings.

In Fig.~\ref{fig:two-emission-test-gg-1} we consider $e^+e^-\to Q \bar Q g_1g_2$, with $\ln m_Q/Q=-6$ and  $\ln {k_{t,1}}/Q=-5$ and $\eta_{1}=1$,  and show the ratio between the exact differential cross-section and the shower approximation for both primary (top row) and secondary (bottom row) emissions. 
The region around the first emission, both for the primary and secondary Lund plane, is mismodelled because double-soft corrections are not included for this test, which are necessary to describe this region properly.%
The shower exactly reproduces the correct QCD expectation in the bulk of phase space, including the region close to the dead-cone boundary, as required by NLL accuracy. 

\begin{figure}
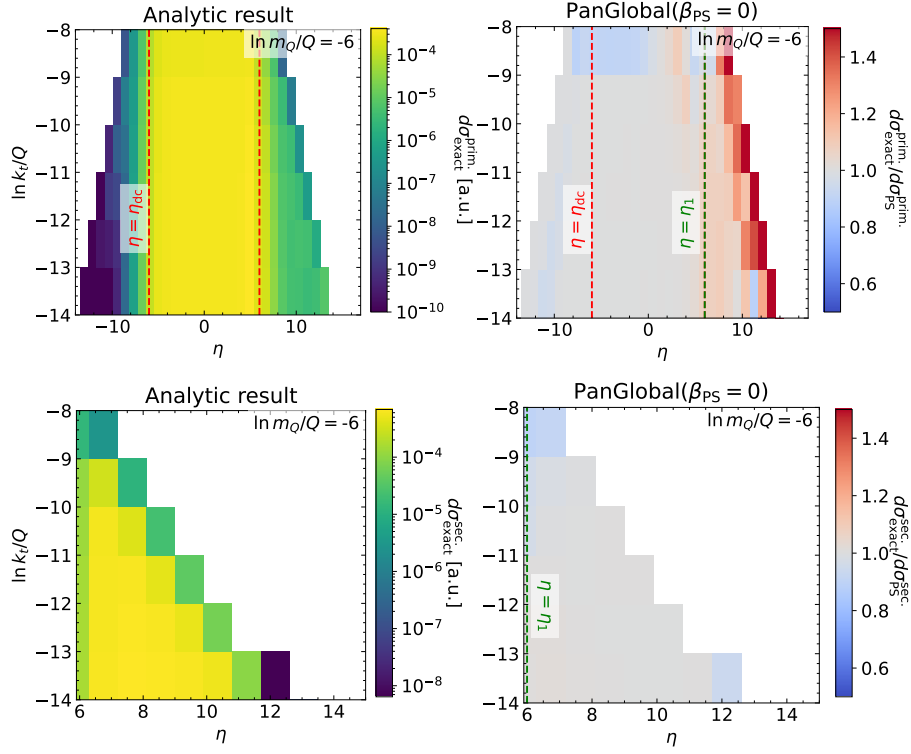

	\centering
  	\includegraphics[width=0.8\textwidth,page=6]{figures/plot_mass_checks_2emsn.pdf}
	\includegraphics[width=0.8\textwidth,page=7]{figures/plot_mass_checks_2emsn.pdf}
\caption{Same as Fig.~\ref{fig:two-emission-test-gg-1}, 
but we request the hardest declustering is generated around  
$(\ln k_{t,1}/Q=-8, \eta_1=6)$.}
	\label{fig:two-emission-test-gg-2}
\end{figure}

In Fig.~\ref{fig:two-emission-test-gg-2} we repeat the same test, but placing the hardest branching at lower energies $\ln k_{t,1}/Q = -8$ and on the dead-cone boundary, i.e.\ around $\eta_{1} = 6$. Note that we only show results for \pg, but that similar conclusions can be drawn for \pl.
It is interesting to observe that the presence of a first gluon at or near the dead-cone region actually enhances the probability of populating the dead-cone region with other gluon emissions (see also Ref.~\cite{Ghira:2025nym}).
We furthermore observe that the area inside the dashed lines, where mass effects may be neglected, is correctly described, as required by NLL accuracy. 
The agreement between the shower and the exact ME is also good for the secondary Lund plane. 
We start seeing deviations from the exact ME when the second gluon is emitted at larger angles than $\eta_{\rm dc}$, even when the second gluon is energy-ordered with respect to the first gluon. 
This particular region contributes to general NLL accuracy, however, we have explicitly verified that the deviation is $N_c$-suppressed.
To cure it would require a modification of the \nods algorithm which we leave for future work. 

\begin{figure}
 	\centering
 	\includegraphics[width=\textwidth,page=1]{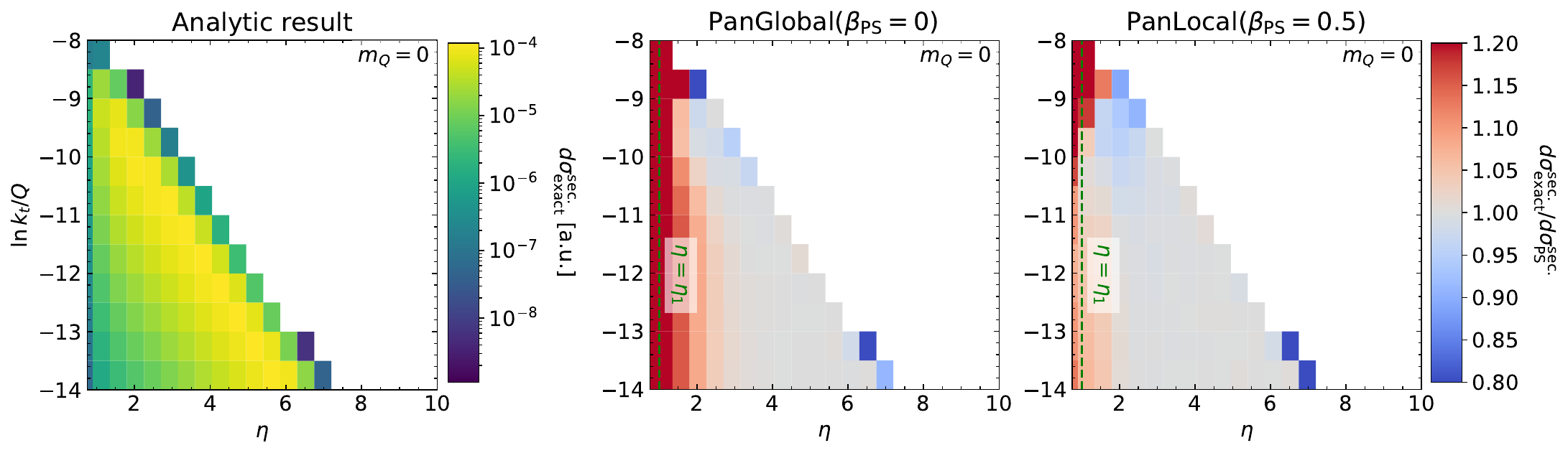}
 	\includegraphics[width=\textwidth,page=4]{figures/plot_mass_checks_2emsn_QQ.pdf}
 	\caption{  
  Ratio between the exact differential cross-section and the shower approximation (\pg, left and \pl, right) for  $e^+e^- \to q\bar q \to q\bar{q}Q\bar{Q}$, as a function of the Lund coordinates $k_t$ and $\eta$ of the softer declustering. We only consider secondary emissions. The first emission is generated around $\ln k_{t,1}/Q = -8$ and $\eta_1 = 1$. 
  The Born quarks are massless, while the heavy quark is massless in the top panel, and massive in the bottom one.  
  }
 	\label{fig:two-emission-test-qq}
 \end{figure}

We conclude our fixed-order tests by considering the $e^+e^- \to q\bar q Q\bar{Q}$ configuration. 
Note that in contrast to our earlier tests, we now require that the Born configuration comprises two massless quarks $q\bar{q}$, in order to focus on the modelling of $g\to Q\bar{Q}$.
In practice, in the generation of the analytic ME, we keep the QCD
contribution (where the $Q\bar{Q}$ pair is connected to a gluon) and
neglect the QED contribution (where the $Q\bar{Q}$ pair comes from the
splitting of a photon),
since the latter configuration is not generated by our shower.
To pass the analysis cuts we require that the two light quarks are not clustered together in a single jet.
The two light quarks are then selected to form the base for the primary Lund plane, and we only consider configurations where the two heavy quarks are clustered together first (i.e.\ secondary Lund plane configurations).
As before, we require the declustering containing the $Q\bar{Q}$ pair to be within a window of $\pm 0.25$ around $\ln k_{t,1}/Q = -8$. 
The result is shown in Fig.~\ref{fig:two-emission-test-qq} for a massless configuration (left) and with $\ln m_Q/Q = -12$ (right). We set $\eta_1 = 1$ in both cases. 
The collinear, large-$\eta$ region is described well for both configurations. 
We see that in the massless case, there is a region close to $\eta\sim \eta_1$ that is not described properly by any of the showers. This region corresponds to soft wide-angle $g\to q\bar{q}$ emissions that starts contributing at NNLL accuracy. 
When considering $m_Q>0$, we observe that the shower weight for $\eta < 4$ does not reproduce the analytic expectation.
Indeed, due to our definition of the ordering variable, when a gluon is emitted at a given transverse momentum $k_{t,1}$,
it cannot lead to a collinear $Q\bar{Q}$ branching if $k_{t,1}<m_Q$.
This is due to the fact that the ``effective'' transverse momentum that governs the parton shower evolution for a $g\to Q \bar{Q}$ splitting is $\kappa_t^2 = k_t^2 + m_Q^2$, and hence the ordering condition $\kappa_{t,2} < \kappa_{t,1}$ implies a second (collinear) branching can only happen for $k_{t,1}=\kappa_{t,1} > m_Q$.
This holds true also when using an ordering variable with $\betaps>0$, as the rapidity of the gluon roughly coincides with that of the $Q$ and the one of the $\bar{Q}$, as we are considering a collinear branching.
However, this missing phase space region does not affect NLL accuracy.
 
\subsection{Tests of spin correlations}
\label{sec:spin-correlations}

\begin{figure}[t!]
	\centering
	\includegraphics[width=0.9\textwidth,page=1]{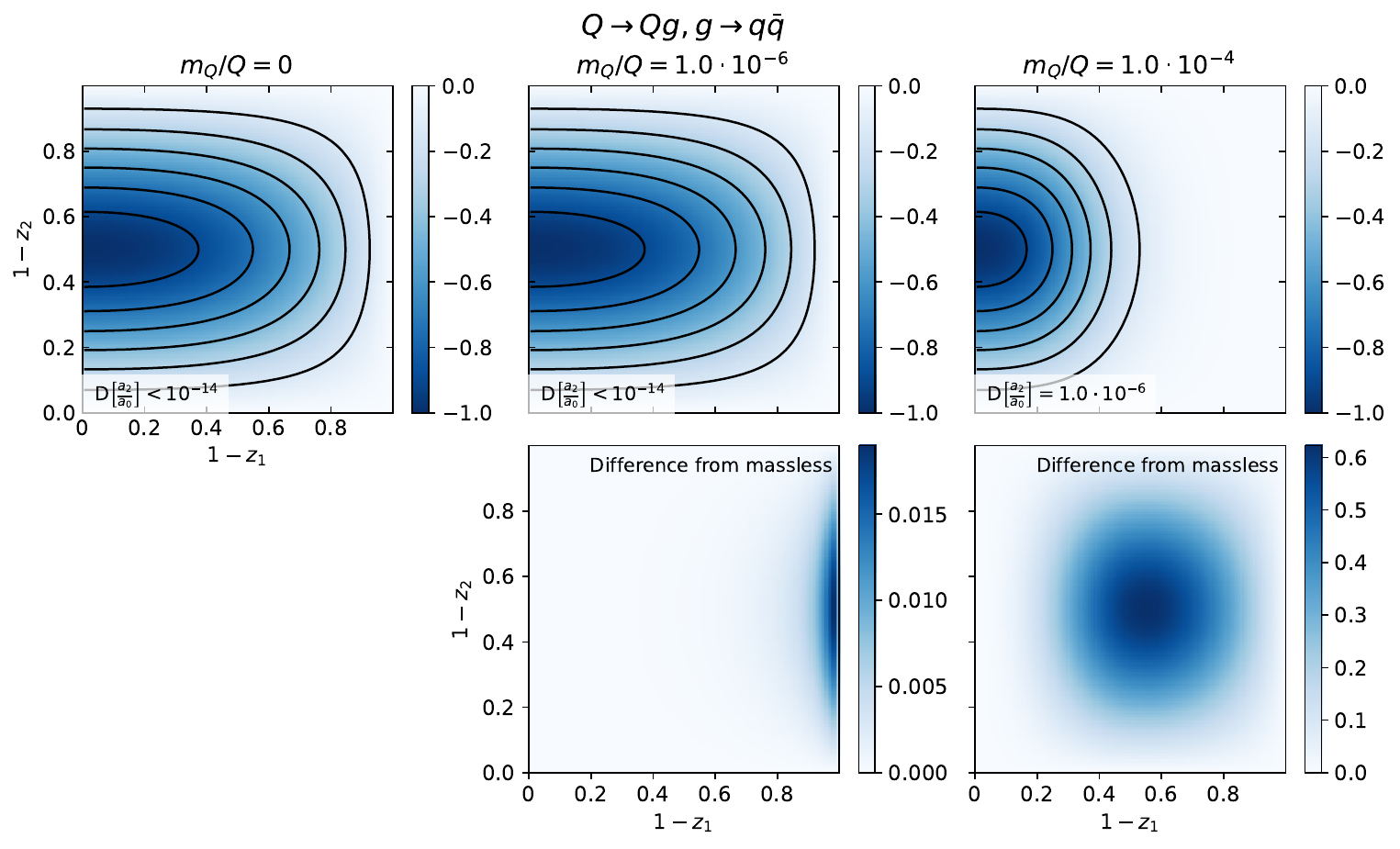}
	\includegraphics[width=0.9\textwidth,page=2]{figures/plot-spin-collinear.pdf}
	\caption{
	The top (bottom) set of plots show a secondary $g\to q\bar{q}$ ( $g\to gg$) splitting, and the columns represent a heavy-quark mass $m_Q/Q$ of $0, 10^{-6}$ and $10^{-4}$ respectively. 
	Within each set, the top panels show the size of the spin correlations $a_2/a_0$ for two collinear splittings, indicated by colour as a function of $z_1$ and $z_2$. Black lines indicate constant values for this ratio. 
	The maximum absolute deviation D[$\frac{a_2}{a_0}$] between
        the shower and the analytic prediction is indicated as well.
	The lower panels show differences to the massless case.
    }
	\label{fig:spin-test}
\end{figure}

\begin{figure}[t!]
	\centering
	\includegraphics[width=0.9\textwidth,page=1]{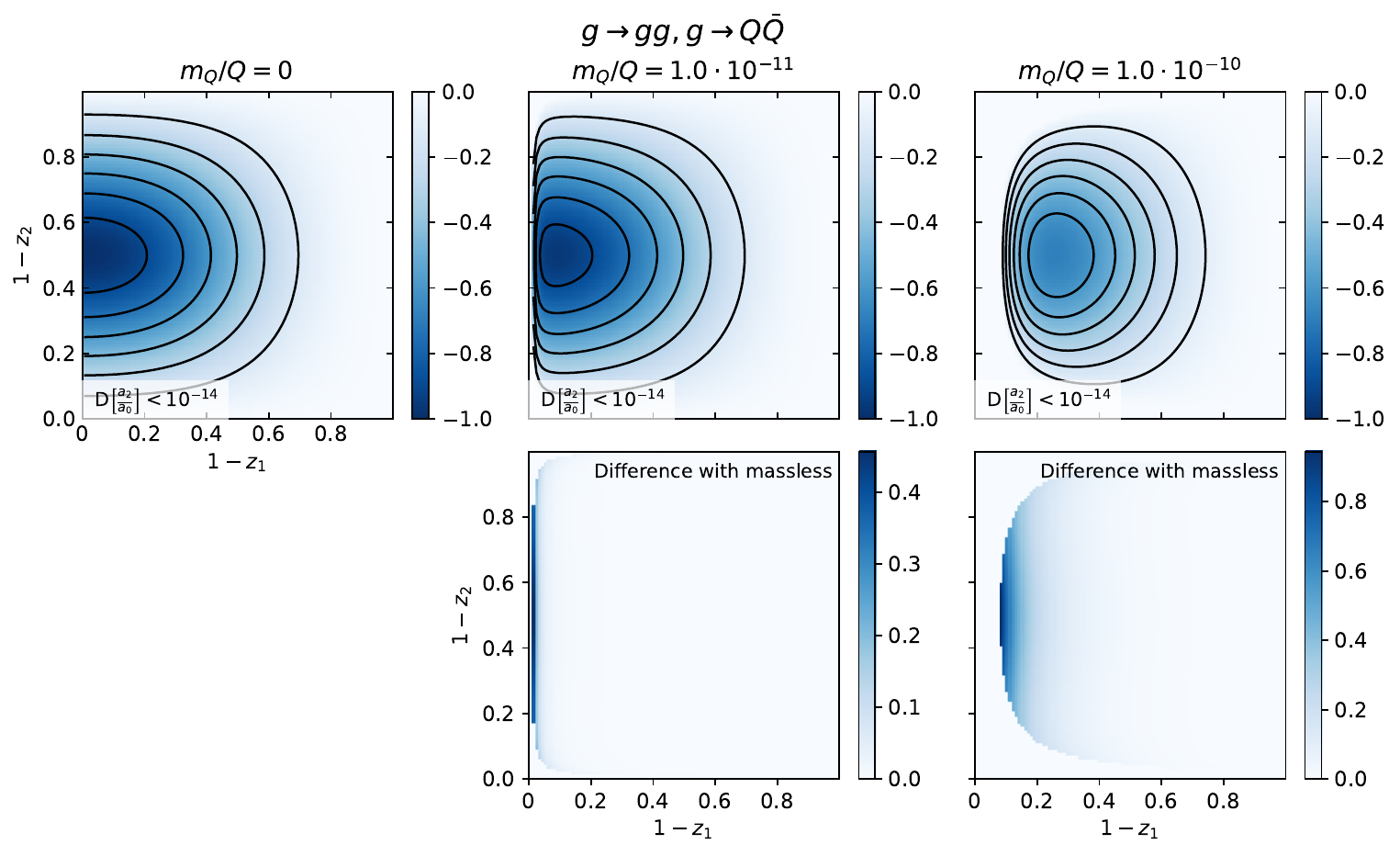}
	\caption{
		Same as Fig.~\ref{fig:spin-test-gg}, but for a $g\to gg$, $g\to Q\bar{Q}$ splitting, and with the columns represent a heavy-quark mass $m_Q/Q$ of $0, 10^{-11}$ and $10^{-10}$ respectively. 
		The sharp transition visible in the two plots in the bottom panel originates from reaching the edge of the collinear $g\to Q\bar{Q}$ shower phase-space.  
	}
	\label{fig:spin-test-gg}
\end{figure}

The presence of masses modifies the strength of the spin correlations, and opens up a third, mass-dependent, helicity channel. 
For example, in massless $q\to qg$ splittings, the emission of a gluon cannot change the helicity of the quark $\lambda$.
When the quark is massive, there is also a non-zero channel allowed where $\lambda \to -\lambda$. 
Similarly, in massless $g\to q\bar{q}$ splittings, quarks are required to have opposite helicities, whereas for heavy quarks equal helicities are also allowed. 
The strength of these new channels are obviously proportional to the mass of the heavy quark. 
We have therefore modified our collinear spin algorithm, introduced in Ref.~\cite{Karlberg:2021kwr}, to take this into account. 

This section aims to provide an $\mathcal{O}(\alpha_s^2)$ validation of our implementation of spin correlations. 
We will only show results for the \pl shower with $
\betaps = 0.5$, as the implementation is identical for the \pg shower and for different $\betaps$ values. 
Furthermore, as we consider two strongly-ordered soft or collinear emissions, the higher-order effects described in Sec.~\ref{sec:log-acc-massless} are negligible and therefore not included in these tests.

In the quasi-collinear limit, cross sections will take a form that is
proportional to
\begin{align}
\frac{d\sigma}{d\Delta \psi_{ij}} \to a_0\left(1 + \frac{a_2}{a_0}\cos(2\Delta \psi_{ij})\right)\,.
\end{align}
Here $\Delta \psi_{ij}$ is the angle between the planes spanned by two
successive branchings, and we sum over all helicities.
A non-zero correlation will be created by the polarisation of an intermediate gluon. 
We will consider a configuration where we start off with a $Q
\to Qg$ splitting, where the gluon carries momentum fraction $1-z_1$, followed by either a collinear $g \to Q\bar{Q}$ or a collinear $g\to gg$, where the $Q$/$g$ of the second splitting caries a momentum fraction $1-z_2$. 
The $a_2/a_0$ ratios for these configurations become (also see Ref.~\cite{Richardson:2018pvo})
\begin{subequations}
\begin{align}
\left(\frac{a_2}{a_0}\right)_{Q\to Qg, g\to Q'\bar{Q}'} &= \frac{2 z_1^2}{z_1(1+z_1^2) + \mu_1 (1-z_1)^3} \times \frac{-2(1-z_2)^2 z_2^2}{\mu_2 + (1 - z_2)z_2(1 - 2(1 - z_2)z_2)}\,, \\
\left(\frac{a_2}{a_0}\right)_{Q\to Qg, g\to gg} &= \frac{2 z_1^2}{z_1(1+z_1^2) + \mu_1 (1-z_1)^3} \times \frac{(1-z_2)^2 z_2^2}{(1-z_2(1-z_2))^2}\,, \\
\label{eq:spin-gg-heavy}
\left(\frac{a_2}{a_0}\right)_{g\to gg, g\to Q'\bar{Q}'} &= \frac{(1-z_1)^2}{(1-z_1 + z_1^2)^2} \times \frac{-2(1-z_2)^2 z_2^2}{\mu_2 + (1 - z_2)z_2(1 - 2(1 - z_2)z_2)}\,.
\end{align}
\end{subequations}
Here $\mu_1$ and $\mu_2$ are quark-mass dependent correction factors, given by
\begin{align}
\mu_1 =  \frac{(1-z_1)m_Q^2}{2 (1-z_1) p_Q\cdot p_g - z_1 m_Q^2}\,, \quad  \mu_2 = \frac{z_2(1-z_2)m_{Q'}^2}{2z_2(1-z_2)p_{Q'}\cdot p_{\bar{Q}'} - m_{Q'}^2 (z_2^2 + (1-z_2)^2)}\,.
\end{align}
It may be checked that for $\mu_1, \mu_2 \to 0$ one recovers the
massless limits quoted in Table~2 of Ref.~\cite{Karlberg:2021kwr}.

The shower-to-analytics comparison of the $a_2/a_0$ ratio as a function of $z_1$ and $z_2$ is shown in Figs.~\ref{fig:spin-test}, where the first splitting is a $Q\to Qg$ one. 
The opening angle for the first splitting is set at $\theta_{Qg} = 10^{-4}$, whereas that for the second splitting is much more collinear with $\theta_{Q'\bar{Q}' / gg} = 10^{-8}$.
We consider a massless benchmark case, together with a quasi-collinear configuration of $m_Q / Q = 10^{-6}$  and a situation where we expect mass effects to significantly alter the spin correlations with $m_Q/Q = 10^{-4}$.
We find perfect agreement with the expected values for $a_2 / a_0$ in each case. 
The spin correlations are maximal for $z_1 \to 1$ and $z_2 \to 0.5$, i.e.\ for the heavy quark carrying most of the momentum fraction in the first splitting, and equal momentum sharing for the second splitting. 
As expected, the presence of a non-zero mass depletes the size of the spin correlations, which is especially visible for the $m_Q/Q = 10^{-4}$ results (rightmost set of plots). 

In Fig.~\ref{fig:spin-test-gg} we show the $a_2/a_0$ ratio for a first splitting being the $g\to gg$ and the second $g\to Q\bar{Q}$. 
We consider the same opening angles as before, but with heavy quark masses of $m_Q/Q = 10^{-11}$ and $m_Q/Q = 10^{-10}$.
We again find perfect agreement between the shower and the analytic expectation (Eq.~\eqref{eq:spin-gg-heavy}). 
The heavy-quark mass again surpresses the amount of spin correlation, in addition to the available shower phase-space for the $g\to Q\bar{Q}$ splitting. 

\section{All order tests of single-logarithmic accuracy}
\label{sec:log-accuracy-tests}
In this section we will discuss three types of all-order tests of our massive showers. 
We will start in Sec.~\ref{sec:lts} with testing whether our shower reproduce the correct NLL corrections relevant for global observables.
In Sec.~\ref{sec:non-global} we test the accuracy of our showers for
an observable which receives non-global logarithms at leading colour,
and we conclude our tests of the all-order accuracy of our showers
with Sec.~\ref{sec:mult} by considering the number multiplicity of subjets produced above a certain transverse momentum $k_{t,{\rm cut}}$.
These tests are designed so as to probe different types of corrections
relevant to claim general NLL/SL accuracy across a range of different
observables. Since we focus on NLL/SL accuracy, we omit the
higher-order corrections discussed in Sec.~\ref{sec:log-acc-massless}.

\subsection{NLL tests for Lund-Tree Shapes}
\label{sec:lts}
Lund-Tree shapes~\cite{Dasgupta:2020fwr,vanBeekveld:2025zjh} are continuously global observables that exploit the Lund-jet-plane representation of QCD radiation. 
We explore both the sum and maximum of the transverse momenta of the declusterings in the primary Lund plane. 
To define the observable, we cluster the entire event into exactly two jets using the Cambridge-Aachen algorithm and then compute 
\begin{align}
  \label{eq:lund-tree-shapes}
	S^{(2)}_b = \sum_{j\in {\rm decl.}} k_{t,j} e^{-b|\eta_j|}\,, \quad 
	M^{(2)}_b = \max_{j\in {\rm decl.}} k_{t,j} e^{-b|\eta_j|}\,, \quad 
\end{align}
with $k_{t,j}$ and $\eta_j$ defined as in Eq.~\eqref{eq:lund-variables}. 

The analytic expressions for the resummation of these observables at NLL accuracy may be obtained using the results of Ref.~\cite{Dhani:2024gtx}. 
The resulting expressions for the cumulative distribution $\Sigma(v)$ for an observable value $v$ are a function of both $\lambda$ and $\lambda_\mu$, defined as 
\begin{align}
\lambda = \alpha_s \ln v/Q \equiv \alpha_s L\,, \quad \lambda_\mu = \alpha_s \ln m_Q/Q \equiv \alpha_s L_\mu\,,
\end{align}
with $m_Q$ the mass of the hard heavy quark legs, and $Q$ the centre-of-mass energy. 
The resummation is valid in the limit where $\lambda$ and
$\lambda_\mu$ are kept fixed and $\alpha_s$ is taken to 0.
In this limit, the resummation is only sensitive to mass effects when $\lambda < \lambda_\mu$.  
To achieve NLL accuracy for these observables, the shower must
correctly predict the dead-cone suppression, which limits the
contribution of soft-gluon radiation in the primary Lund plane (this
effect would be present for $\lambda<(1+b)\lambda_\mu$), in
addition to describing correctly the number-of-light-flavours change
in the running coupling and $K_{\rm CMW}$, which affects the shower's
Sudakov. Note that the dead-cone suppression and the mass thresholds
in the running coupling already have an impact at the LL accuracy. 

From the shower perspective, we isolate the NLL terms by fixing values
of $\lambda$ and $\lambda_\mu$ and running several decreasing values
of $\alpha_s$. 
We fix $\lambda=-0.5,\lambda_\mu=-0.2 $ and compute the following
ratio
\begin{align}
  \label{eq:nll-ratio-lund-tree}
	\frac{\Sigma_{\rm PS}(\alpha_s,\lambda, \lambda_\mu) - \Sigma_{\rm NLL}(\alpha_s,\lambda, \lambda_\mu)}{\Sigma_{\rm NLL}(\alpha_s,\lambda, \lambda_\mu)}\,.
\end{align}
NLL accuracy requires that the above ratio goes to zero in the $\alpha_s \to 0$ limit. 
In practice, we take the limit $\alpha_s\to 0$ by performing a polynomial
extrapolation.
For our central value of the extrapolation we use $\alpha_s \in [0.002, 0.005,0.008,0.01]$.
The error is then calculated by adding the statistical uncertainty plus the difference in the extrapolated value using another set of $\alpha_s$ values ($\alpha_s \in [0.0025,0.004,0.008]$) in quadrature. 
Subleading-$N_c$ corrections have been included using the \nods method~\cite{Hamilton:2020rcu}. 

\begin{figure}[t!]
 	\centering
 	\includegraphics[width=0.7\textwidth,page=2]{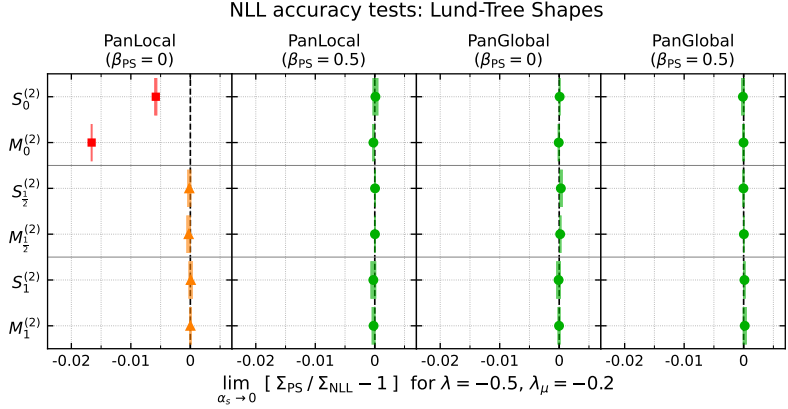}
 	\caption{Summary of deviations from NLL for several global observables for $\lambda = -0.5$ and $\lambda_\mu = -0.2$.}
	\label{fig:nll-tests-global-obs}
\end{figure}

In Fig.~\ref{fig:nll-tests-global-obs} we evaluate Eq.~\eqref{eq:nll-ratio-lund-tree} in the $\alpha_s \to 0$ limit for the observables $M_b$ and $S_b$ using the \pl and \pg showers. 
For \pg, we expect NLL agreement for any value $\betaps < 1$. 
Contrary, for \pl, we need $0 < \betaps < 1$,
whereas $\betaps = 0$ suffers from a similar transverse recoil issue
as the one discussed for the Dire-v1 shower in
Sec.~\ref{sec:contours}.
We decide to show \pl with $\betaps = 0$ nonetheless, because it gives
an indication of the size of NLL failures due to this kinematic issue. 
Furthermore, we show results for $b = 0, 0.5, 1$, and $\betaps = 0, 0.5$.
We
observe agreement for all combinations of showers and observables except for \pl with $\betaps = 0$ when computing $b=0$ Lund-Tree shapes, as expected. 

\begin{figure}[t!]
 	\centering
 	\includegraphics[width=0.49\textwidth,page=1]{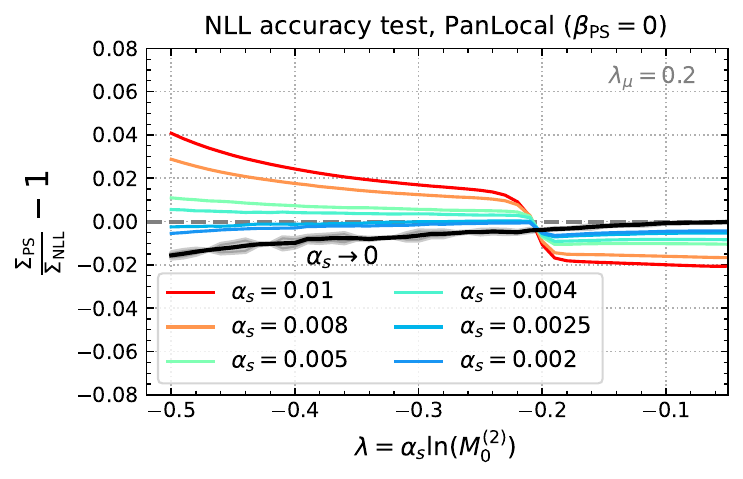}
        \includegraphics[width=0.49\textwidth,page=3]{figures/plot-dist-paper.pdf}
 	\includegraphics[width=0.49\textwidth,page=5]{figures/plot-dist-paper.pdf}
        \includegraphics[width=0.49\textwidth,page=7]{figures/plot-dist-paper.pdf}
 	\caption{Evaluation of Eq.~\eqref{eq:nll-ratio-lund-tree}
          for the \pl shower with $\betaps=0$ (LL, left) and
          $\betaps=0.5$ (NLL, right) as a function of $\alpha_s \ln
          (M^{(2)}_{0})$ (left) and $\alpha_s \ln
          (M^{(2)}_{1/2})$ (right), for different values of $\alpha_s$. The NLL
          expectation corresponds to the grey, dashed line. The black
          line corresponds to the $\alpha_s\to 0$
          extrapolation.}
	\label{fig:nll-tests-global-obs-dist}
\end{figure}

It is also illustrative to study the convergence of the cumulative distribution of $\lambda$ to the NLL result as a function of $\alpha_s$. 
This is shown in Fig.~\ref{fig:nll-tests-global-obs-dist}, where we again
evaluate Eq.~\eqref{eq:nll-ratio-lund-tree}
as a function of $\lambda=\alpha_s \ln (M^{(2)}_{0})$ and $\lambda=\alpha_s \ln (M^{(2)}_{1/2})$ for different values of
$\alpha_s$. Results are shown for both the \pl and \pg shower with $\betaps = 0$ and $\betaps =
0.5$.
We fix $\lambda_\mu = -0.2$. 
This limit is obtained by doing an extrapolation of the $\alpha_s\in [0.0025,0.004, 0.01]$ values, and adding again as an error the statistical uncertainty and the difference with doing the extrapolation with $\alpha_s \in [0.002, 0.004,0.005]$ in quadrature. 
We observe that for $\betaps = 0.5$ the ratio indeed goes to zero as $\alpha_s \to 0$, while for $\betaps = 0$ the ratio does not go to zero, as expected from the fact that this shower is not NLL accurate. 

\begin{figure}[t!]
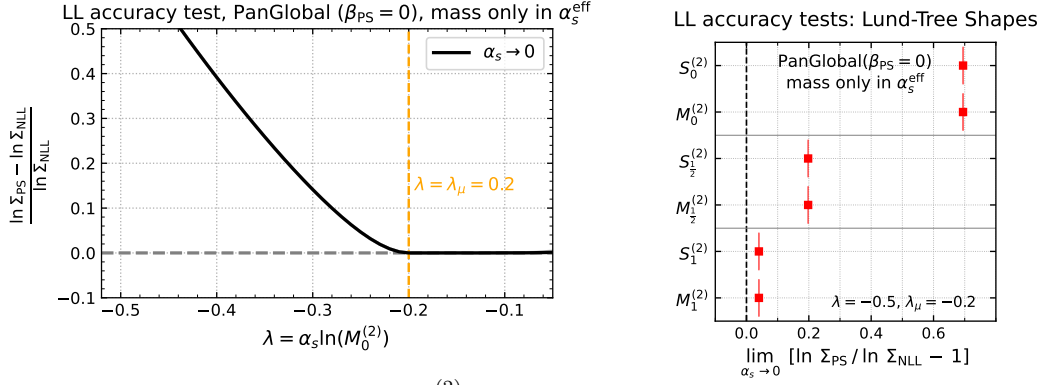

  \centering
  \begin{subfigure}[c]{0.53\textwidth}
    \centering
    \includegraphics[width=\linewidth,page=9]{figures/plot-dist-paper.pdf}
    \caption{$\alpha_s\to 0$ extrapolation of the $M_0^{(2)}$
      observable. The LL expectation corresponds to the grey, dashed
      line while the orange line indicates the $\lambda =
      \lambda_\mu$.}
    \label{fig:ll-tests-global-obs-dist-a}
  \end{subfigure}
  \hspace{0.5cm}
  \begin{subfigure}[c]{0.34\textwidth}
    \centering
    \includegraphics[width=\linewidth,page=4]{figures/plot-summary.pdf}
    \caption{LL test for all considered global observables.}
    \label{fig:ll-tests-global-obs-dist-b}
  \end{subfigure}
  \caption{Leading-logarithm (LL) accuracy tests for the \pg shower with all mass effects turned off except in the running of the coupling.}
  \label{fig:ll-tests-global-obs-dist}
\end{figure}

At this stage it is important to highlight that without mass
corrections in the shower kinematics, the showers would not even achieve LL accuracy for global observables once the scale of the observable falls below the particle mass scale. 
This is due to the fact that the dead cone at LL effectively acts as a $\Theta$-function, vetoing radiation emitted at angles smaller than the dead-cone angle relative to the emitting particle.
This effect can be seen in Fig.~\ref{fig:ll-tests-global-obs-dist}. 
In Fig.~\ref{fig:ll-tests-global-obs-dist-a}, we again consider the
$M^{(2)}_0$ observable, but now performing the logarithmic accuracy
test at LL.
This is done by evaluating the ratio
\begin{align}
	\label{eq:ll-test}
	\frac{\ln \Sigma_{\rm PS}(\alpha_s,\lambda, \lambda_\mu) - \ln \Sigma_{\rm NLL}(\alpha_s,\lambda, \lambda_\mu)}{\ln \Sigma_{\rm NLL}(\alpha_s,\lambda, \lambda_\mu)}\,.
\end{align}
LL accuracy requires that this ratio goes to zero in the $\alpha_s \to 0$ limit. 
We consider the \pg shower with mass thresholds in the effective coupling, but using massless kinematics in the shower evolution and in the shower matrix elements. 
A clear failure of the LL test is visible for $\lambda < \lambda_{\mu}$, that is, when the scale of the observable drops below the mass scale. 
The effect is increasingly large as $\lambda_\mu-\lambda$ increases.
Fig.~\ref{fig:ll-tests-global-obs-dist-b} shows the
LL discrepancy for a series of observables, focusing on $\lambda=-0.5$.
The size of the violation depends on the observable parameter
$b$.
This is primarily due to the fact that these LL violations would in
general be present for $\lambda<(1+b)\lambda_\mu$. 
This means that the phase space region that is incorrectly described by a massless shower
is larger for observables with $b=0$ than for observables with $b=1$.  

One may wonder what happens if the dead-cone region is removed artificially, i.e.\ by a phase-space veto, but without consistently incorporating mass effects into the splitting functions or shower kinematics. 
We investigate this possibility in App.~\ref{app:simple-veto}, where we introduce a simplified shower model that mimics heavy-quark effects by imposing a veto on the radiation around the dead-cone angle. 
More concretely, we veto $Q\to Qg$ splittings if $\eta_g<\eta_{\rm dc} + \eta_s$ and $g\to Q\bar Q$ splittings if $k_t < m_Q$. 
Choosing $\eta_s = 0$ would be sufficient to reproduce LL results, but as discussed in Appendix~\ref{app:simple-veto}, this choice would lead to the wrong NLL prediction. 
Instead, adopting $\eta_s = -\frac{1}{2}$ would achieve NLL accuracy
for global event shapes.\footnote{This value ensures that we also
  reproduce the right single-logarithmic term arising from the integration
  of the massive eikonal factor in the Sudakov (see
  Appendix~\ref{app:simple-veto} for details).}
Moreover, while this simplified prescription would give the correct NLL prediction, it would fail for non-global observables. 
In non-global observables, logarithmically enhanced contributions arise from cascades where gluons emit soft gluons into a restricted region of phase space. 
Capturing these effects requires the shower to reproduce the full
structure of the massive eikonal factor --- and, in particular, its
exact rapidity dependence --- for nested soft emissions.
A simple angular veto around the dead cone only reproduces the total
rate integrated over the rapidity of the emission but fails to capture
the differential distribution. As a result, such a shower cannot
reproduce the non-global logarithms associated with heavy-quark
radiation.

\subsection{Non-global energy flow}
\label{sec:non-global}
We now move on to testing the accuracy of our showers for an observable which receives non-global logarithms at leading colour. 
Namely, we consider the energy flow into a rapidity slice of width $\Delta$,
\begin{align}
S_{\Delta} = \sum_{i} E_i \Theta(|y_i| - \Delta)\,,
\end{align}
where $E$ and $y$ are the energy and rapidity of each particle, respectively. 
At the resummation level, there are potentially two kinds of logarithms that arise when considering massive quarks. 
These are single-logarithmic terms of the form $\lambda = \alpha_s L$
with $L = \ln Q/S_{\Delta}$ and $\lambda_\mu = \alpha_s L_\mu$ with
$L_\mu = \ln Q/m_Q$, with $\ln Q$ the logarithm of the hard scale of
the process.
For our tests of non-global logarithms, we are interested in the
kinematic configuration where the total energy in the slice is small,
$S_{\Delta}\ll Q$, while we keep the width of the slice finite,
$\Delta\sim 1$.
In this context, we also want to keep the ratio $m_Q/Q\sim 1$, meaning
that the dead cone and the edge of the slice are at commensurate
rapidities, so as to have non-trivial single-logarithmic effects from
the mass with respect to the massless case.
In summary, this means that we consider the limit $\alpha_s\ll 1$,
$\lambda\sim 1$, and $\Delta\sim L_\mu\sim 1$, i.e.\ $\lambda_\mu\to 0$.
In this regime, the presence of the dead cone will lower the total
emission rate of primary gluons in the region in and around the slice. 

\begin{figure}[t!]
\centering
\includegraphics[width=0.48\textwidth,page=1]{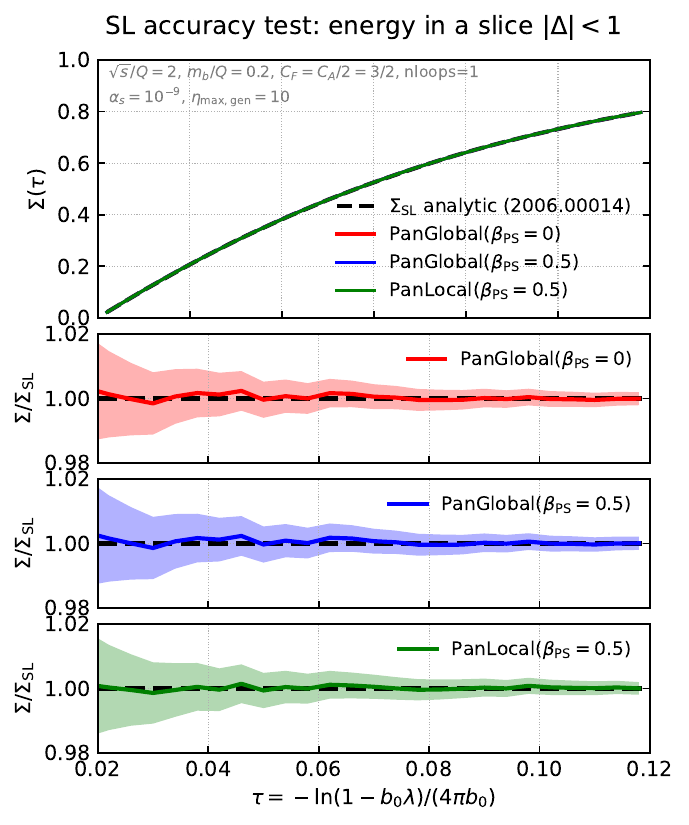}
\includegraphics[width=0.48\textwidth,page=2]{figures/plot-rapidity-slice.pdf}
\caption{Cumulative cross-section (Eq.~\eqref{eq:cum-slice}) for the
  energy in a rapidity window of $|\Delta|=1$ (left) and $|\Delta|=2$
  (right), as a function of the evolution time defined in
  Eq.~\eqref{eq:tau-def}. The three bottom panels show the ratio
  between each shower and the analytic results.}
\label{fig:rapidity-slice}
\end{figure}

Our reference resummation for this observable is obtained from the code developed in Ref.~\cite{Balsiger:2020ogy}. 
In this reference resummation, the so-called evolution time $\tau$ is defined as
\begin{align}
  \label{eq:tau-def}
\tau = \int_0^L \dd\ell \frac{\alpha_s(\ell)}{2\pi}\,,  
\end{align}
with $\alpha_s(\ell)$ being the one-loop running coupling. 
In the massless case, Eq.~\eqref{eq:tau-def} can be integrated to give 
\begin{align}
\tau \Big\vert_{m_Q=0}= -\frac{\ln \left(1 -2 b^{n_f=5}_0 \lambda\right)}{4\pi b^{n_f=5}_0}\,,
\end{align}
with $b^{n_f}_0$ given by Eq.~\eqref{eq:beta0-def}. 
In the massive case, we need to account for flavour thresholds in the coupling. 
The relevant expression for $\tau$ then becomes
\begin{align}
  \label{eq:tau-def-massive}
\tau \Big\vert_{m_Q\neq 0} = \int_0^{L_\mu} \frac{\dd\ell}{2\pi} \frac{\alpha_s}{1-2\alpha_s b^{n_f=5}_0\ell} + \int_{L_\mu}^L \frac{\dd\ell}{2\pi} \frac{\alpha_s}{1-2\alpha_s b^{n_f=5}_0 L_\mu-2\alpha_s b^{n_f=4}_0(\ell-L_\mu)}\,. 
\end{align}
In the regime where $m_Q\sim Q$, $\lambda_\mu\sim 0$, and Eq.~\eqref{eq:tau-def-massive} reduces to
\begin{align}
  \label{eq:tau-def-massive-final}
\tau \Big\vert_{\frac{m_Q}{Q}\sim\mathcal{O}(1)}= - \frac{\ln \left(1 -2 b^{n_f=4}_0 \lambda\right)}{4\pi b^{n_f=4}_0}. 
\end{align}
To test the shower accuracy, we assume a quark mass of $m_Q/Q = 0.2$, i.e.\ the dead cone lies around $\eta_{\rm dc} = 1.6$, with $Q$ the scale used to define $\alpha_s$. 
The centre-of-mass energy, $\sqrt{s}$, of the back-to-back dipole is set to $2Q$. We use $C_F = C_A/2 = 3/2$ as in Ref.~\cite{Balsiger:2020ogy}.
We consider two configurations for the rapidity-slice window: $\Delta = 1$, so that the dead cone lies fully in the vetoed region, and $\Delta = 2$, where the dead cone lies partially in the allowed region. 
We then calculate the cumulative distribution of $\tau$, defined as 
\begin{align}
	\label{eq:cum-slice}
	\Sigma(\tau) = \int_0^{\tau} d\tau' \frac{d \sigma(\tau')}{d\tau'}\,,	
\end{align}
with $\tau = 0$ marking $\lambda = 0$, i.e.\ the case where $S_{\Delta} = Q$. 
Our canonical $\lambda = 0.5$ value corresponds to $\tau \sim 0.12-0.13$. 
The results are obtained with an asymptotically small value of $\alpha_s = 10^{-9}$, using the techniques developed in Ref.~\cite{Hamilton:2021dyz}.
We show the results for the two values of $|\Delta|$ in Fig.~\ref{fig:rapidity-slice}, for $0.02 < \tau < 0.12$ for \pg with $\betaps = 0,0.5$ and \pl with $\betaps = 0.5$, demonstrating perfect agreement across the full range of $\tau$. 
The statistical uncertainty on the reference calculation dominates the uncertainty band. 

\begin{figure}[t!]
	\centering
	\includegraphics[width=0.48\textwidth,page=3]{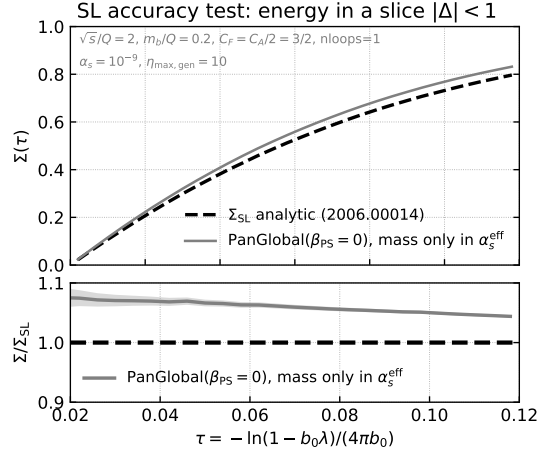}
	\caption{Same as Fig.~\ref{fig:rapidity-slice} (left), but for a shower where the mass effects are partially turned off. See text for more details.  }
	\label{fig:rap-slice-massless}
\end{figure}

To quantify the impact of mass corrections, we show in Fig.~\ref{fig:rap-slice-massless} the result we obtain when we keep the flavour thresholds in $\alpha_s^{\rm eff}$, but turn off mass corrections in the splitting functions and in the shower kinematics.
Effectively, this has the effect of removing the dead-cone suppression. 
Note that for the chosen value of $\Delta = 1 < \eta_{\rm dc}$, the region of phase space where primary emissions of gluons off the massive Born particles is suppressed since the dead-cone lies outside of the vetoed region. 
Yet, we observe up to 10\% deviations with respect to the correct SL
result.
This mismatch is due to the fact that, already at
$\mathcal{O}(\alpha_s)$, $\Sigma(\tau)$ will be sensitive to the mass
effects in the soft eikonal factor, i.e.\ be influenced by the
presence of the dead cone. The massive eikonal factor would result in
a further suppression of the radiation in the slice, hence a reduction
of $\Sigma(\tau)$ as seen in the plot.
We have explicitly checked that an analytic integration of the massive
eikonal factor agrees with the observed deviation in
Fig.~\ref{fig:rap-slice-massless}.

\subsection{Lund subjet multiplicity}
\label{sec:mult}
We conclude our logarithmic accuracy tests by considering the number of subjets produced above a certain transverse momentum $k_{t,{\rm cut}}$. 
Algorithmically, this is obtained by reclustering the event with the Cambridge-Aachen algorithm, and counting the number of declusterings that have a transverse momentum above $k_{t,{\rm cut}}$. 
The resummation of this observable follows a different logarithmic structure than that of global observables. 
In fact, the subjet multiplicity has a double-logarithmic resummation structure, $\alpha_s^nL^{2n}$ with $L = \ln k_{t,{\rm cut}}/Q$. 
When accounting for mass effects, a second logarithm, $L_\mu = \ln m_Q/Q$, must be taken into account. 
We work in the regime in which both $L$ and $L_\mu$ are large and negative, and we perform the resummation up to next-to-double-logarithmic (NDL) accuracy. 
This means that we organise the resummation in terms of $\xi = \alpha_s L^2$ and $\xi_\mu = \alpha_s L_{\mu}^2$ using the following expansion
\begin{align}
\langle N(\alpha_s,L,L_\mu)\rangle = h_1(\xi, \xi_\mu) + \sqrt{\alpha_s}h_2(\xi, \xi_\mu) + \mathcal{O}(\rm NNDL)\,,
\end{align}
and compute up to the $h_2$ function. 
Mass effects are only relevant when $k_{t,\rm cut} < m_Q$.
In this regime, the resummation is modified by, for instance, the presence of the dead cone, which suppresses the emission of gluons that would otherwise be able to produce subjets above $k_{t,{\rm cut}}$. 
The analytic computation is detailed in App.~\ref{app:analytics-multiplicity}. 

 \begin{figure}[t!]
 	\centering
 	\includegraphics[width=0.48\textwidth,page=3]{figures/plot-mult-summary.pdf}
 	\includegraphics[width=0.48\textwidth,page=1]{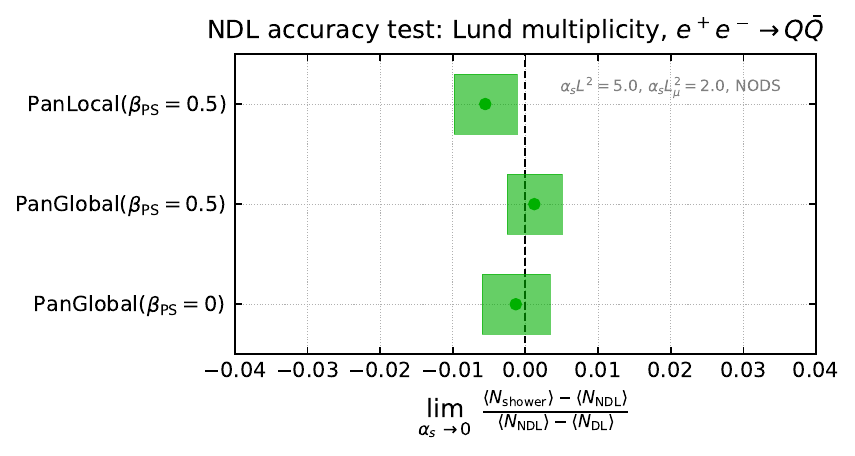}
 	\caption{The result of Eq.~\eqref{eq:mult-test-ndl} for three different showers and two Born-level processes: light-quark initiated jets (left) and heavy-quark initiated jets (right). Green points indicate that the result is consistent with zero within $2\sigma$.}
 	\label{fig:multiplicity}
 \end{figure}
To test the NDL terms produced by the shower, we compute the following ratio
\begin{align}
  \label{eq:mult-test-ndl}
\frac{N_{\rm PS} - N_{\rm NDL}}{N_{\rm NDL}-N_{\rm DL}}\,,
\end{align}
which should vanish in the $\alpha_s \to 0$ limit if the shower is correct at NDL accuracy. 
Similar to what was done in the test of global event shapes, we run the shower at different values of $\lbrace \alpha_s, k_{t,{\rm cut}}, m_Q\rbrace$ while keeping $\xi = 5$ and $\xi_\mu = 2$ fixed. 
Here, we choose to run the shower with the values $\ln k_{t,{\rm
    cut}}/Q \in [-31.25, -62.5, -125, -250, -500, -1000]$ and perform
the extrapolation using the subset $\ln k_{t,{\rm cut}}/Q \in [-31.25,
-62.5, -125,-1000]$. 
The uncertainty on the extrapolation is given by adding the statistical error and the difference between performing the extrapolation with an alternate set of $\ln k_{t,{\rm cut}}/Q \in [-31.25,  -125, -250,  -1000]$.
The result is shown in Fig.~\ref{fig:multiplicity} for two different hard processes, $e^-e^+ \to q\bar{q}$ and  $e^-e^+ \to Q\bar{Q}$.
We observe that all showers are consistent with the full-colour NDL expectation.

\section{Preliminary phenomenological results and comparison to data}
\label{sec:pheno}
In this section we present phenomenological results obtained with our new NLL showers. 
To this end we interface the \panscales code~\cite{vanBeekveld:2023ivn} with \pythia (v8.317)~\cite{Bierlich:2022pfr} for hadronisation. 
For the \pl shower, we use the default values for the Monash tune~\cite{Skands:2014pea}, except for $\alpha_s$, which we keep at its canonical value of $\alpha_s(m_Z) = 0.118$. 
For \pg, we use the $\beta_{\rm PS}$-dependent tunes of Ref.~\cite{vanBeekveld:2024wws}. 
We have explicitly verified by running our \pg showers with the Monash tune (and $\alpha_s(m_Z) = 0.118$) that the tuning effects do not affect the physics discussion below.
The values of the $c$ and $b$ masses are set to $m_c = 1.5$~GeV, $m_b = 4.8$~GeV. 
The handling of flavour thresholds in the coupling is discussed in Sec.~\ref{sec:kcmw-threshold}, and we 
include higher order effects in the massless limit, as explained in Sec.~\ref{sec:log-acc-massless}. 

\begin{figure}[t!]
\centering
\begin{subfigure}{0.49\textwidth}
  \centering
  \includegraphics[page=1,width=\textwidth]{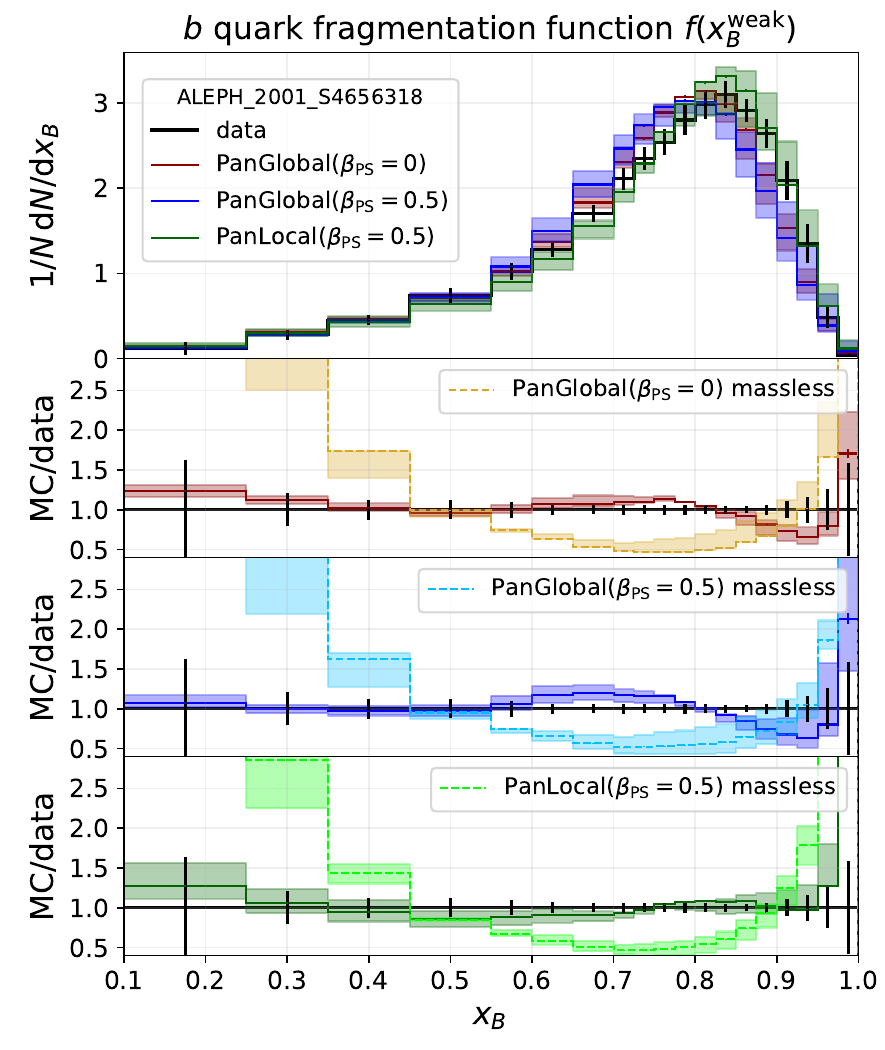}
  \caption{Fragmentation function for weakly decaying $b$ quarks by ALEPH~\cite{ALEPH:2001pfo}. Rivet analysis: \href{https://rivet.hepforge.org/analyses/ALEPH_2001_I558327.html}{ALEPH\_2001\_I558327} by P.~Richardson.}
  \label{fig:brag}
\end{subfigure}
\hfill
\begin{subfigure}{0.49\textwidth}
  \centering
  \includegraphics[page=8,width=\textwidth]{figures/rivet_plots.pdf}
  \caption{Total jet broadening for $b$ events at the $Z$ peak by L3~\cite{L3:2004cdh}. Rivet analysis: \href{https://rivet.hepforge.org/analyses/L3_2004_I652683.html}{L3\_2004\_I652683} by A.~Jueid.}
  \label{fig:bbroad}
\end{subfigure}
\caption{Comparison between \panscales{} showers with mass effects, namely \pg, with $\betaps=0$~(red) and $\betaps=0.5$~(blue), as well as  \pl with $\betaps=0.5$~(green), with LEP measurements at the $Z$ peak (data, black) for the $b$-quark fragmentation function~\ref{fig:brag} and the total jet broadening in $b$ events~\ref{fig:bbroad}.
In the bottom panels, ratio to data for the massive and massless (dashed line, brighter colour) showers are illustrated.
All showers include hard matrix-element corrections~\cite{Hamilton:2023dwb} for the first emission in the massless limit, while retaining the unmatched massive shower matrix element around the dead cone.
Both \pg shower variants contain double soft~\cite{FerrarioRavasio:2023kyg} and NNLL Sudakov effects~\cite{vanBeekveld:2024wws} with a dynamical number of active flavours.
Error bars indicate statistical uncertainty, filled bands represent scale variation uncertainty.
}
\label{fig:pheno}
\end{figure}

In Fig.~\ref{fig:pheno}
we show the $b$-quark fragmentation function (\ref{fig:brag})
and the total jet broadening for $b$-flavoured events (\ref{fig:bbroad}).
In the bottom panels we display the ratio to the data, also for the corresponding massless shower.
The analyses are performed with Rivet~\cite{Bierlich:2019rhm}.

We first consider the $b$-quark fragmentation function $f$ as a function of $x_B^{\rm weak}$, which is defined as the energy of the weakly-decaying $B$ meson ($B\to l\nu D^{(*)}$) normalised to the beam energy.
The non-perturbative part of the fragmentation function itself depends explicitly on the quark mass, but the perturbative mass of the $b$-quark actually also plays a predictive role, in that it shifts the average value of $x_B^{\rm weak}$, defined as
\begin{align}
\langle x_B^{\rm weak} \rangle = \int dx_B^{\rm weak} \,  x_B^{\rm weak} f(x_B^{\rm weak})\,,
\end{align} 
to higher values (see e.g.\ Ref.~\cite{PhysRevD.27.105}). 
Because massive quarks radiate less than massless ones, a massive quark will retain more of its energy before it undergoes hadronisation compared to a massless quark. 
The inclusion of mass corrections in the perturbative evolution of the shower is therefore crucial to get the correct shape for this observable, as is also clearly observed in the left-hand side of Fig.~\ref{fig:brag}. 
As seen in the bottom three panels of this figure, massless showers completely fail to predict the correct the shape of this observable.
This is caused by the simple fact that the bottom quark is treated as
massless in that case, which will then push $\langle x_B^{\rm weak}
\rangle$ to lower values.
%
%
For our showers, we see values of $\langle x_B^{\rm weak}
\rangle$ that are roughly 20\% smaller in the massless shower compared to the
massive one. 
However, after the inclusion of mass effects we do correctly predict the shape of the distribution.
Note that we did not perform a dedicated tuning exercise. 
It can therefore be expected that the data agreement can be improved
after tuning.

For the total jet broadening observable with $b$-flavoured events, i.e.\ Fig.~\ref{fig:bbroad}, we notice good agreement with the data for both the massless and massive variants in the majority of the probed phase space. 
Being a $k_t$-like observable (like $M^{(2)}_0$ or $S^{(2)}_0$), the broadening gains sensitivity to the quark mass around $B_T = m_b/Q$ (indicated by the vertical red dashed line). 
There, radiation will be suppressed with respect to the massless case.
It is clearly visible that around  $B_T = m_b/Q$ , the inclusion of mass
effects impacts the observable, bringing the massive shower in better
agreement with the data than the massless counterpart. Note however
that hadronisation effects may be significant in this region.

\section{Conclusions}
\label{sec:conclusions}
In this work we have presented the first final-state parton showers that demonstrably account for the NLL logarithmic terms associated with non-zero quark masses.
We have designed two classes of showers, the \pl (with $\betaps=1/2$) and \pg showers ($\betaps=0$ and $\betaps=1/2$). The \pl shower uses a dipole-local recoil scheme, whereas the \pg showers use an event-wide recoil scheme. In the massless limit, the newly introduced shower algorithms reduce to those presented in Refs.~\cite{Dasgupta:2020fwr, Hamilton:2023dwb,FerrarioRavasio:2023kyg, vanBeekveld:2024wws}. To correctly describe mass corrections at NLL accuracy, we have relied on three main ingredients: (i) kinematic maps based on light-like reference vectors whose recoil assignment respects NLL principles, (ii) an emission probability that simplifies to the massive eikonal in the soft limit, and to the massive DGLAP splitting function in the quasi-collinear limit, and (iii) a scheme to account for flavour thresholds in the running of the effective coupling. Our showers also incorporate (quasi-)collinear spin correlations and subleading-colour corrections.

We have performed a battery of tests to validate the logarithmic accuracy of our showers. To validate both the kinematic maps and the emission probability we have carried out matrix-element tests up to $\mathcal{O}(\alpha^2_s)$. The showers reproduce the exact matrix element in the bulk of phase space as well as the transition towards the dead-cone region. In the deep-collinear regime, we have observed deviations from the analytic benchmark in the form of power-suppressed contributions which, however, do not spoil NLL accuracy. To probe the nested structure of the showers we have examined a series of observables with distinct resummation structures at NLL accuracy. The heavy-quark mass introduces a new scale and corresponding logarithm that must be resummed to all orders. We have presented shower-to-resummation comparisons for Lund-Tree shapes, non-global energy flow and Lund subjet multiplicity. For the two classes of observables we used the results from Refs.~\cite{Dhani:2024gtx} and \cite{Balsiger:2020ogy}, respectively, while for the Lund subjet multiplicity we computed the resummation baseline. In all cases, we have found that the \panscales showers reproduce the resummation benchmark, thus rigorously validating their logarithmic accuracy.

NLL accuracy with massive quarks is a key step towards phenomenological applications of our showers. In this paper we have shown some first shower-to-data comparisons for the $b$-quark fragmentation function and the broadening for $b$ events at LEP energies. Mass corrections are crucial to correctly capture the shape of the $b$-quark fragmentation function, whereas the broadening distribution is only affected at values $B_T \lesssim m_b/Q$.  

The resulting showers are released in the public \panscales framework, and will simultaneously account for NLL/SL corrections whenever the mass scale is of relevance, while preserving the correct NNLL/NSL corrections for (non)global event-shapes in the massless limit, guaranteed by a careful matching of the two physical domains. 
A natural continuation of this work is to implement higher-order ingredients with full mass effects. 
Besides this, we also plan to extend these algorithms to initial-state radiation so that our showers can be used to fully exploit the heavy-flavour programme at proton-proton and electron-proton colliders. 

\section*{Acknowledgements}
We are grateful to our PanScales collaborators (Sam Bates, Mrinal Dasgupta, Basem El-Menoufi, Keith Hamilton, Jack Heliwell, Petr Jakubcik, Alexander Karlberg, Matvei Kuzmin, Marco Leitão, Pier Monni, Gavin Salam, Nicolas Schalch, Ludovic Scyboz, and Silvia Zanoli) for their work on the code, the underlying philosophy of the approach and comments on this manuscript. We also wish to thank Gavin Salam for many insightful discussions during the course of this project.
This research was supported
by the Italian Ministry of Universities and Research (MUR) under the
FIS grant (CUP:D53C24005480001, FLAME) (SFR),
by the Ramón y Cajal program under grant RYC2022-037846-I, ERDF (grant
PID2024-161668NB-100) and by grant NSF PHY-2309135 to the Kavli
Institute for Theoretical Physics (KITP) (ASO),
by the Dutch Research Council (NWO) under project number
VI.Veni.232.190 (MvB),
by the European Research Council (ERC) under the European Union’s
Horizon 2020 research and innovation programme (grant agreement
No. 788223, PanScales),
and by the French Agence Nationale de la Recherche (grant
ANR-25-CE31-7192, P2S2) (GS). 

\appendix
\section{Mapping coefficients}
\label{app:mapping-coefficients}
\subsection{\pl mapping coefficients}
\label{app:pl-mapping}
Here we collect the mapping coefficients for \pl (Eq.~\eqref{eq:pl-map}).
Solving the momentum conservation and on-shell constraints we find 
\begin{subequations}\label{eq:pl-map-coefficients}
\begin{align}
  a_i &= \frac{\bar \mu + \mu_t + \mu_i - \mu_j + \sqrt{\lambda(\bar \mu, \mu_t+\mu_i, \mu_j)}}{2\widetilde\kappa_{ij}(1- \widetilde\kappa_{ij} \beta_k)}\, , \\
  a_j &= \frac{\bar \mu - \mu_t - \mu_i + \mu_j - \sqrt{\lambda(\bar \mu, \mu_t+\mu_i, \mu_j)}}{2\widetilde\kappa_{ij}(1- \widetilde\kappa_{ij} \beta_k)} \, ,\\
  b_i &= \frac{\bar \mu + \mu_t + \mu_i - \mu_j - \sqrt{\lambda(\bar \mu, \mu_t+\mu_i, \mu_j)}}{2\widetilde\kappa_{ij}(1- \widetilde\kappa_{ij} \alpha_k)} \, ,\\
  b_j &= \frac{\bar \mu - \mu_t - \mu_i + \mu_j + \sqrt{\lambda(\bar \mu, \mu_t+\mu_i, \mu_j)}}{2\widetilde\kappa_{ij}(1- \widetilde\kappa_{ij} \alpha_k)},
\end{align}
\end{subequations}
with $\lambda$ defined in Eq.~\eqref{eq:kallen}, and where
\begin{align}
  \mu_t & = \widetilde\kappa^2_{ij} \alpha_k \beta_k - \mu_k\,, \quad
  \bar \mu = (1- \widetilde\kappa_{ij} \alpha_k)(1-\widetilde\kappa_{ij} \beta_k)\,, \quad
  \mu_t = \frac{k_t^2}{\dipm^2}\,.
\end{align}
The normalised masses $\mu_{i,j}$ are defined using the post-branching masses, i.e.\ $\mu_i = m_i^2 /\dipm^2$. 

\subsection{\pg mapping coefficients}
\label{app:pg-mapping}
Here we report the expressions for the coefficients entering the \pg map of Eq.~\eqref{eq:pgmap}.
Before doing so, we introduce the four-vector
\begin{equation}
p_\perp^{\mu} = p_t\left(\sin\varphi \,\hat{n}_{\perp,1}^\mu+\cos\varphi \,\hat{n}_{\perp,2}^\mu\right), \qquad p_t^2 =2 \alpha_k \beta_k \widetilde{s}_{ij} -m_k^2=\kappa_t^2-m_k^2,
\end{equation}
where $\widetilde{s}_{ij}=2\widetilde{n}_i \cdot \widetilde{n}_j$, and $(\hat{n}_{\perp,1}^\mu$, $\hat{n}_{\perp,2}^\mu$) are two orthogonal unit vectors transverse to the plane defined by $\widetilde{n}_i$ and $\widetilde{n}_j$. Note that $p_\perp^\mu$ and $k_\perp^\mu$, which appears in Eq.~\eqref{eq:pgmap}, are related via a simple rescaling:
\begin{equation}
  k_\perp^\mu = (1+\delta \widetilde{\kappa}_{ij})\, p_\perp^\mu.
\end{equation} 
We will also use the invariants $\widetilde{s}_{i}=2\widetilde{n}_i \cdot Q$, and
$\widetilde{s}_{j}=2\widetilde{n}_j \cdot Q$.

As stated in the main text, the factors $a_k$ and $b_k$ are given by
\begin{equation}
 a_k = \alpha_k (1+ \delta\, \widetilde{\kappa}_{ij})r_k, \qquad 
 b_k = \beta_k (1+ \delta \,\widetilde{\kappa}_{ij})r_k\,.
\end{equation}
The rescaling coefficients $\delta$ and $r_k$ are obtained after
imposing that the invariant mass of the event is conserved, in
addition to imposing the on-shell condition for $p_k$. After some
algebra, one finds
\begin{equation}
  \label{eq:pg-rescaling}
\delta=\frac{2 (p_{t}^2+2 p_\perp \cdot Q)}{\widetilde{s}_i+\widetilde{s}_j-2 \widetilde{\kappa}_{ij} (p_{t}^2+p_\perp \cdot Q)+\sqrt{\left(\widetilde{s}_i+\widetilde{s}_j-2 \widetilde{\kappa}_{ij} (p_{t}^2+p_\perp \cdot Q)\right)^2+4 (p_{t}^2+2 p_t \cdot Q) \left(\widetilde{s}_{ij}-\widetilde{\kappa}_{ij}^2 p_{t}^2\right)}},
\end{equation}
and
\begin{equation}
	\label{eq:pg-rk}
  r_k = \sqrt{1-\frac{\delta\,\widetilde{\kappa}_{ij}\, m_k^2(\delta \, \widetilde{\kappa}_{ij}+2)}{\kappa_t^2(1+ \delta \,\widetilde{\kappa}_{ij})^2}}.
\end{equation}
The map coefficients for $\bar{p}_i^\mu$ and $\bar{p}_j^\mu$ read
\begin{subequations}
	\label{eq:pg-mapping}
\begin{align}
a_i & =\bar{d}_i \left(\frac{1}{\widetilde{\kappa}_{ij}}+\delta -a_k \right), &b_i =& (1-\bar{d}_j) \left(\frac{1}{\widetilde{\kappa}_{ij}}+\delta -b_k \right), \\
a_j & = (1-\bar{d}_i) \left(\frac{1}{\widetilde{\kappa}_{ij}}+\delta -a_k \right), & b_j =& \bar{d}_j \left(\frac{1}{\widetilde{\kappa}_{ij}}+\delta -b_k \right), 
\end{align}
\end{subequations}
where 
  \begin{equation}
    \bar{d}_i  = \frac{1}{2}\left(1+\kappa_{ij}+\mu_i-\mu_j\right)\,, \qquad
    \bar{d}_j  = \frac{1}{2}\left(1+\kappa_{ij}+\mu_j-\mu_i\right)\,, 
\end{equation}
and
\begin{subequations}
  \begin{align}
      \mu_{i,j} & = \frac{m^2_{i,j}}{m_{ij}^2}\\
    \kappa_{i,j}^2 & = 1-2(\mu_i+\mu_j) + (\mu_i-\mu_j)^2\\
    m_{ij}^2 & = \left(\frac{1}{\widetilde{\kappa}_{ij}}+\delta-a_k\right)\left(\frac{1}{\widetilde{\kappa}_{ij}}+\delta-b_k\right)\widetilde{s}_{ij}.
\end{align}
\end{subequations}
Note that, unlike the massless case, the map coefficients $a_j$ and $b_i$ are different from zero.

\section{Lund multiplicity with heavy-quarks at NDL}
\label{app:analytics-multiplicity}
Here we report the analytic calculation of the Lund multiplicity up to next-to-double logarithmic accuracy (NDL). For a detailed 
derivation in the massless case see Section 3.2 in Ref.~\cite{Medves:2022ccw}. Before entering into details of the calculation let us first
clarify the meaning of NDL accuracy in the presence of three disparate scales: $Q$, the centre of mass energy; $\ktcut$ a transverse 
momentum threshold imposed on each subjet pair; and $m$, the heavy-quark mass.
We define two logarithms, $L\equiv \ln(Q/\ktcut)$ and $L_{\mu} \equiv \ln(Q/m)$, and work in the limit in which both $L$ and $L_{\mu}$ are large. Then, 
achieving double-logarithmic accuracy implies resumming terms of order $\alpha_s^n (\xi_1\xi_2)^n$, where $\xi_i$ is either $L$ or $L_\mu$. For instance, 
when $n=1$ we must account for all terms that involve $L^2, L L_\mu$ and $L_\mu^2$. Similarly, at next-to-double logarithmic accuracy we must capture terms of order $\alpha_s^n (\xi_1\xi_2)^{n-1}$. Finally, since we are interested in computing mass corrections to the Lund multiplicity, we assume throughout the calculation that $L<L_\mu$ or, equivalently, $\ktcut \ll m$. 

We first discuss the double-logarithmic result for the Lund multiplicity in one hemisphere that we denote $N^{(\DL)}_{i}$, with $i = q,g,Q$ corresponding to a light-quark-, gluon- or heavy-quark-initiated jet.  At DL accuracy we need to consider the emission of soft-and-collinear gluons. For $i=q,g$, the double-log branching probability is unmodified by the presence of a mass threshold and we thus recover the massless results, i.e.
\begin{equation}
\label{eq:nlight-dla}
N^{(\DL)}_{i\in\lbrace q,g\rbrace} =  1 + \frac{C_i}{C_A} (\cosh\nu-1)\,,\\
\end{equation}
where we introduced 
\begin{equation}
  \nu = \sqrt {\bar\alpha L^2} = \sqrt{\frac{2\alpha_s C_A L^2}{\pi}}.
\end{equation}
For a heavy-quark initiated jet we need to account for the fact that the initial gluon cannot be radiated at rapidities larger than $L_\mu$. The all-orders expression then reads
\begin{align}
\label{eq:nQ-dla-integral}
N^{(\DL)}_Q &=  1 +   \frac{2\as C_F}{\pi}\int_0^L \dd\ell \int_0^{\min(\ell, L_\mu)} \dd\eta N_g(L-\ell)\, , \\
& = 1 + \frac{C_F}{C_A} [\cosh \nu - \cosh(\nu-\nu_\mu)], 
\end{align}
where
\begin{equation}
 \nu_\mu=\sqrt{\bar\alpha L_\mu^2} =  \sqrt{\frac{2\alpha_s C_A L_\mu^2}{\pi}}.
 \end{equation}
The distribution of Lund declusterings at a given $k_t$ can be obtained by differentiating the $N^{(\DL)}_i$ results with respect to $L$, i.e.
 \begin{align}
\label{eq:nQ-dla}
n^{(\DL)}_{i\in\lbrace q,g\rbrace} &=  \frac{C_i}{C_A} \sqrt{\bar \alpha} \sinh \nu\, , \\
n^{(\DL)}_Q &=  \frac{C_F}{C_A} \sqrt{\bar \alpha}[ \sinh \nu -  \sinh(\nu-\nu_\mu)] .
\end{align}
Note that the second term in $n^{(\DL)}_Q$ is absent when the scale is below the mass threshold. 

At NDL accuracy we need to consider three types of mass corrections: (i) mass thresholds in the running of the coupling, (ii) hard-collinear branchings, and (iii) phase-space effects in the dead cone boundary. The NDL correction thus reads
\begin{equation}
\delta N^{(\NDL)}_i = \delta N^{(\NDL)}_{i,\beta_0} +  \delta N^{(\NDL)}_{i,{\rm hc}} + \delta N^{(\NDL)}_{i,{\rm rc}}\delta(i-Q)\, .
\end{equation}
The last term only affects heavy-quark initiated jets while the other two corrections are present for the three Born process under consideration. We now proceed to discuss the computation of each correction independently.
\paragraph{Running coupling corrections.} We first consider the NDL correction associated with the one-loop running of the strong coupling including mass-thresholds. That is, we need to account for the fact that one of the emissions in the chain (either the first or any secondary one) will be get a correction due to the fact that above the mass-threshold there are $5$ active flavours, while below it we have to switch to $4$ in a continuous way. This is achieved via    
\begin{align}
\label{eq:n-ndl-rc}
\as(\ell) & \approx \as + 2\as^2 [\beta^{(5)}_0 \min(\ell ,L_\mu) + \beta^{(4)}_0 \max(0,\ell-L_\mu)] + \mathcal{O}(\as^3)\, ,
\end{align}
with $\beta_0^{(n_f)}=(11 C_A- 2 n_f)/(12\pi)$ and $n_f$ being the number of active flavours. 
For a light-quark/gluon-initiated jet we find that the running coupling correction to the Lund multiplicity reads
\begin{align}
\label{eq:nlight-ndl-rc-step1}
\delta N^{(\NDL)}_{i\in\lbrace q,g\rbrace,\beta_0} & = \frac{4\alpha_s^2 C_i}{\pi} \left[ \int_0^L \dd \ell_1 [\delta(\ell_1) + n^{(\DL)}_{g}(\ell_1)] \right. \, \nonumber \\
&\left. \times \int_{\ell_1}^L \dd \ell_2 [\beta^{(5)}_0  \min(\ell_2, L_\mu) +  \beta^{(4)}_0  \max(0,\ell_2-L_\mu)] (\ell_2-\ell_1)  N^{(\DL)}_g(L-\ell_2)\right]\, ,
 \end{align}
where the $\delta(\ell_1)$ term accounts for the correction on the primary emission. Inserting the corresponding expressions for $N^{(\DL)}_g$ and $n^{(\DL)}_g$ we can perform all integrals analytically and obtain
\begin{align}
\label{eq:nlight-ndl-rc}
\delta N^{(\NDL)}_{i\in\lbrace q,g\rbrace,\beta_0} & = \frac{C_i}{C^2_A} \sqrt{\bar\alpha} \frac{\pi}{4}\Big\lbrace \beta_0^{(5)} [\nu_\mu \cosh\nu + \nu_\mu(2 \nu-\nu_\mu) \sinh\nu - \sinh\nu_\mu \cosh(\nu-\nu_\mu)]\nonumber \\
      &+ \beta_0^{(4)} [(\nu-\nu_\mu) \cosh\nu + (\nu-\nu_\mu)^2 \sinh \nu - \cosh\nu_\mu\sinh(\nu-\nu_\mu)]\Big\rbrace.
\end{align}
For a heavy-quark-initiated jet we need to include both the dead-cone restriction on the rapidity of the first emissions and the $\beta_0$-correction, thus we modify Eq.~\eqref{eq:nlight-ndl-rc-step1} accordingly, i.e. 
\begin{align}
\label{eq:nQ-ndl-rc-step1}
\delta N^{(\NDL)}_{Q,\beta_0} & = \frac{4\alpha_s^2 C_F}{\pi} \Big\lbrace \int_0^L \dd \ell [\beta_0^{(5)} \min(\ell,L_\mu) + \beta_0^{(4)} \max(0,\ell-L_\mu)] \min(\ell,L_\mu) N^{(\DL)}_g(L-\ell) \,  \\
&+ \int_0^L \dd \ell_1n^{(\DL)}_{Q}(\ell_1) \int_{\ell_1}^L \dd \ell_2  [\beta_0^{(5)} \min(\ell_2,L_\mu) + \beta_0^{(4)} \max(0,\ell_2-L_\mu)] (\ell_2-\ell_1)  N^{(\DL)}_g(L-\ell_2) \Big\rbrace \, \nonumber.
 \end{align}
 Computing the integrals, we find the $\beta_0$-correction to heavy-quark initiated jets  
 \begin{align}
\label{eq:nheavy-ndl-rc}
\delta N^{(\NDL)}_{Q,\beta_0} & = \frac{C_F}{C^2_A} \sqrt{\bar\alpha} \frac{\pi}{4}\Big\lbrace \beta_0^{(5)} [\nu_\mu \cosh\nu + \nu_\mu(2 \nu-\nu_\mu) \sinh\nu - 2 \nu_\mu (\nu-\nu_\mu) \sinh(\nu-\nu_\mu) \nonumber \\ 
& -\sinh\nu_\mu \cosh(\nu-\nu_\mu)] + \beta_0^{(4)} [(\nu-\nu_\mu) (\cosh\nu-\cosh(\nu-\nu_\mu)) + (\nu-\nu_\mu)^2 \nonumber \\
& \times (\sinh\nu-\sinh(\nu-\nu_\mu)) - (\cosh\nu_\mu-1) \sinh(\nu-\nu_\mu)]\Big\rbrace.
\end{align}

\paragraph{Hard-collinear corrections.} In the quasi-collinear regime, finite-energy corrections become relevant at NDL accuracy. We thus need to compute the correction associated with the integral of the non-divergent part of the splitting functions which is standard to denote as $B$ coefficients. At NDL, we can compute the $B$ coefficients using the massless splitting function, since mass corrections enter as power-suppressed terms. However, when considering a gluon splitting into a quark-anti-quark pair, the number of active flavours has to be updated when crossing the mass threshold. The $B$-coefficients thus read
\begin{subequations}\label{eq:bcoeff}
\begin{align}
B_q &\equiv \int_0^1 \dd z \Big( \frac{P_{q\to qg}(z)}{2C_F}  - \frac{1}{z}\Big) = -\frac{3}{4}, & B_g &\equiv  \int_0^1 \dd z \Big( \frac{P_{g\to gg}(z)}{2C_A}  - \frac{1}{z} \Big)= - \frac{11}{12} \\
B_{gQ}& \equiv \int_0^1 \dd z \frac{P_{g\to q\bar q}(z)}{2C_A} \Big\vert_{n_f=1}= \frac{T_R}{3 C_A}, & B_{gq}& \equiv n_f B_{gQ} \,
\end{align}
\end{subequations}
where $B_{gQ}$ corresponds to a $g\to Q\bar Q$ splitting above the heavy-quark mass threshold, while $B_{gq}$ represents $g\to q\bar q$ splittings with the number of light flavours set to $n_f=4$. We also need to take into account that the hard-collinear splitting can be either primary or secondary; flavour-diagonal or flavour-changing.
Taking all this into account, we write the NDL hard-collinear correction to a gluon-initiated jet as
 \begin{align}
\label{eq:ng-ndl-hc-step1}
\delta N^{(\NDL)}_{g,\rm{hc}} & =  \bar\alpha \Big\lbrace \int_0^L \dd \ell_1
  \left[B_g \delta(\ell_1) + B_g n_g^{(\DL)}(\ell_1)\right]  
  \int_{\ell_1}^L
  \dd \ell_2\, N^{(\DL)}_g(L-\ell_2) \nonumber \\
  & + \int_0^{L_\mu} \dd\ell B_{gQ} [2N^{(\DL)}_Q(L-\ell) - N^{(\DL)}_g(L-\ell) ] \nonumber \\
& + \int_0^L\dd\ell B_{gq} [2N^{(\DL)}_q(L-\ell) - N^{(\DL)}_g(L-\ell)] \nonumber \\
& +  \int_0^{L_\mu} \dd\ell_1 n^{(\DL)}_g(\ell_1) \int_{\ell_1}^{L_\mu}\dd \ell_2 B_{gQ} [2N^{(\DL)}_Q(L-\ell_2) - N^{(\DL)}_g(L-\ell_2)] \nonumber \\
&+ \int_0^{L} \dd\ell_1 n^{(\DL)}_g(\ell_1)\int_{\ell_1}^{L}\dd \ell_2 B_{gq} [2N^{(\DL)}_q(L-\ell_2) - N^{(\DL)}_g(L-\ell_2)] \Big\rbrace \,,
\end{align}
 where we have used the shorthand notation
 \begin{equation}
N^{(\DL)}_Q (\lambda) = 1 + \frac{C_F}{C_A} [\cosh(\sqrt{\bar \alpha}\lambda) -\cosh(\sqrt{\bar \alpha}(L-L_\mu))]\,.
 \end{equation}
 The first line in Eq.~\eqref{eq:ng-ndl-hc-step1} corresponds to a flavour-diagonal splitting for which the result is identical to the massless case. The remaining four terms describe a hard-collinear $g\to Q\bar Q$ splitting that occurs either in the primary branch (second and third lines) or in a secondary branch (fourth-to-sixth lines). The limits of integration have been split to account for momentum scales above or below the mass threshold. Once again, we can perform the integrations and get 
 \begin{align}
 \label{eq:ng-ndl-hc}
\delta N^{(\NDL)}_{g,\rm{hc}} & = \sqrt{\bar\alpha} \Big\lbrace   \frac{B_g}{2}  (\nu \cosh\nu + \sinh\nu)
      + B_{gq} \Big[\Big(\frac{C_F}{C_A}-\frac{1}{2}\Big) \nu  \cosh\nu +  \Big(\frac{3}{2}-\frac{C_F}{C_A}\Big) \sinh\nu\Big]\nonumber \\
      &+ B_{gQ}\Big [\Big(\frac{C_F}{C_A}-\frac{1}{2}\Big)\nu_\mu \cosh \nu - \Big(\frac{C_F}{C_A}+\frac{1}{2}\Big) \cosh(\nu-\nu_\mu) \sinh \nu_\mu + 2 \sinh \nu_\mu\Big] \Big\rbrace\, .
 \end{align}

For a light-quark-initiated jet, we can write the hard-collinear correction as 
\begin{align}
\label{eq:nq-ndl-hc-step1}
\delta N^{(\NDL)}_{q,\rm{hc}} & = \frac{C_F}{C_A} \bar\alpha \Big\lbrace \int_0^L \dd \ell_1
  \left[B_q \delta(\ell_1) + B_g n_g^{(\DL)}(\ell_1)\right]  
  \int_{\ell_1}^L
  \dd \ell_2\, N^{(\DL)}_g(L-\ell_2) \nonumber \\
  &+ \int_0^{L_\mu} \dd\ell_1 n^{(\DL)}_g(\ell_1) \int_{\ell_1}^{L_\mu}\dd \ell_2 B_{gQ} [2N^{(\DL)}_Q(L-\ell_2) - N^{(\DL)}_g(L-\ell_2)] \nonumber \\
&+\int_{0}^L \dd\ell_1 n^{(\DL)}_g(\ell_1) \int_{\ell_1}^{L}\dd \ell_2 B_{gq} [2N^{(\DL)}_q(L-\ell_2) - N^{(\DL)}_g(L-\ell_2)] \Big\rbrace \,.
\end{align}
In this case, the mass correction comes from a secondary branching in which a gluon splits into a $Q\bar Q$ (second and third lines). Integrating the previous expression we find
 \begin{align}
\label{eq:nq-ndl-hc}
\delta N^{(\NDL)}_{q,\rm{hc}} & = \frac{C_F}{C_A} \sqrt{\bar\alpha} \Big\lbrace  B_q\sinh\nu + \frac{B_g}{2}  (\nu \cosh\nu - \sinh\nu) \nonumber \\
& + B_{gq}\Big[\frac{c_\delta}{2}\nu\cosh\nu + \Big(1-\frac{3 c_\delta}{2}\Big)\sinh\nu -(1-c_\delta)\nu \Big] \nonumber \\
& + B_{gQ}\Big[c_\delta[\sinh(\nu-\nu_\mu)-\sinh\nu] +\frac{c_\delta}{2} \nu_\mu(\cosh\nu +\cosh(\nu-\nu_\mu)) \nonumber \\
& + (2 - (1+\frac{c_\delta}{2})\cosh(\nu-\nu_\mu)) (\sinh\nu_\mu - \nu_\mu) \Big]  \Big\rbrace \, ,
\end{align}
where $c_\delta\equiv 2C_F/C_A-1$.

Lastly, for a heavy-quark initiated jet we obtain
 \begin{align}
\label{eq:nQ-ndl-hc-step1}
\delta N^{(\NDL)}_{Q,\rm{hc}} & = \bar\alpha\frac{C_F}{C_A}  \Big\lbrace \int_0^{L_\mu} \dd \ell B_q N_g^{(\DL)}(L-\ell)
 + \int_{0}^L \dd \ell_1 n^{(\DL)}_Q(\ell_1)  \int_{\ell_1}^L \dd \ell_2\, N^{(\DL)}_g(L-\ell_2) \nonumber \\
 & +  \int_{0}^{L_\mu} \dd \ell_1 n^{(\DL)}_Q(\ell_1)  \int_{\ell_1}^{L_\mu} \dd \ell_2 B_{gQ}  [2N^{(\DL)}_Q(L-\ell_2) - N^{(\DL)}_g(L-\ell_2)]  \nonumber \\
 & + \int_{0}^L \dd \ell_1 n^{(\DL)}_Q(\ell_1)  \int_{L_\mu}^L \dd \ell_2 B_{gq}  [2N^{(\DL)}_q(L-\ell_2) - N^{(\DL)}_g(L-\ell_2)]  \Big\rbrace \,.
\end{align}
In the previous expression, the first two terms correspond to the hard-collinear correction being on the primary branch or secondary and flavour diagonal, respectively. Lines two and three encapsulate a hard-collinear gluon splitting into a $Q\bar Q$ pair in some secondary branch. The final result for this Born flavour reads
\begin{align}
\label{eq:nQ-ndl-hc}
\delta N^{(\NDL)}_{Q,\rm{hc}} & = \frac{C_F}{C_A}\sqrt{\bar\alpha} \Big \lbrace 
          B_q  [\sinh\nu - \sinh(\nu-\nu_\mu)]\nonumber \\
        & + \frac{B_g}{2} \Big [\nu \cosh\nu - (\nu-\nu_\mu) \cosh(\nu-\nu_\mu) + \sinh(\nu-\nu_\mu) - \sinh\nu\Big] \nonumber \\
        & + B_{gQ} \Big[ c_\delta [\frac{\nu_\mu}{2} (\cosh\nu+\cosh(\nu-\nu_\mu)) - \frac{1}{2}(\sinh\nu_\mu-\nu_\mu) \cosh(\nu-\nu_\mu) - \sinh\nu + \sinh(\nu-\nu_\mu)] \nonumber \\
         &       + (2-\cosh(\nu-\nu_\mu)) (\sinh\nu_\mu-\nu_\mu) \Big]\nonumber \\
        & + B_{gq} \Big[ \frac{c_\delta}{2} [\nu(\cosh\nu-\cosh(\nu-\nu_\mu)) + \nu_\mu (2+\cosh(\nu-\nu_\mu)) - 3 (\sinh\nu-\sinh(\nu-\nu_\mu))]\nonumber \\
          &        + \sinh\nu-\sinh(\nu-\nu_\mu) - \nu_\mu\Big]  \Big \rbrace \,.
\end{align}

\paragraph{Phase-space correction at dead-cone boundary.} In the DL calculation, we assumed that the dead cone boundary acted as a Heaviside function when integrating soft-and-collinear emissions over rapidity. At NDL, we need to refine such phase-space restriction to account for the fact emissions at rapidities larger than the dead-cone angle are not strictly forbidden, but instead (exponentially) suppressed. To do so, we integrate the eikonal form factor to radiate a soft gluon up to NDL accuracy
\begin{equation}
  \label{eq:dc-ps-correction}
\int_0^\ell \dd\eta \frac{1}{[1+ e^{2(\eta-L_\mu)}]^2} =  \ell \Theta(L_\mu-\ell) + (L_\mu +\lambda_{\rm dc}) \Theta(\ell -L_\mu) + \mathcal{O}(\NNDL).
\end{equation}
In the previous expression, the first term corresponds to the massless result. The second term accounts for mass corrections, with the factor $\lambda_{\rm dc}=-1/2$ being the NDL correction. Plugging this correction to the matrix element into the all-orders expression for the Lund multiplicity of a heavy-quark initiated jet yields
\begin{align}
\delta N^{(\NDL)}_{Q,{\rm dc}} & =  \frac{2\as C_F}{\pi}\lambda_{\rm dc} \int_{L_\mu}^L \dd\ell N_g(L-\ell), \nonumber \\
&= \frac{C_F}{C_A} \sqrt{\bar\alpha} \lambda_{\rm dc} \sinh(\nu-\nu_\mu)\, .
\end{align}

\section{An alternative approach to achieve partial NLL accuracy}
\label{app:simple-veto}
In this appendix we explore an alternative approach to achieving NLL accuracy for massive quarks. 
The underlying idea is to keep a massless version of both the momentum map (we use  the \pl map) and the splitting functions. Then, mass corrections enter the shower algorithm through (i) the variable-flavour number effective running coupling
(including $K_{\rm CMW}$) and (ii) a veto of emissions that mimics the dead-cone effect.
We show here that this simple approach is sufficient to give the correct NLL result for the LTS observables (Sec.~\ref{sec:lts}), but it fails to capture the full pattern of soft radiation at single-logarithmic accuracy, and in particular it does not give the correct result for non-global observables.

The starting point of the construction is the standard massless soft eikonal for radiation (with momentum $k$) from a $q\bar q$ dipole,
\begin{equation}
  \label{eq:massless-eikonal}
	W_{q\bar q}^{(0)}(k) = \frac{2p_q\cdot p_{\bar q}}{(p_q\cdot k)(p_{\bar q}\cdot k)} \,.
\end{equation}
For massive quarks, the soft eikonal instead reads
\begin{equation}
	\label{eq:massive-eikonal}
	W_{Q\bar Q}^{(m)}(k)
	=
	\frac{2p_Q\cdot p_{\bar Q}}{(p_Q\cdot k)(p_{\bar Q}\cdot k)} +  \frac{m^2}{(p_Q\cdot k)^2} + 
	\frac{m^2}{(p_{\bar Q}\cdot k)^2} \, .
\end{equation}
The additional terms in Eq.~\eqref{eq:massive-eikonal} compared to Eq.~\eqref{eq:massless-eikonal} suppress collinear radiation off each massive leg at scales
\begin{equation}
	\eta \lesssim \eta_{\rm dc} \equiv \ln \left(\frac{2E_Q}{m}\right).
\end{equation}
Our aim is to reproduce the integrated effect of Eq.~\eqref{eq:massive-eikonal} while still using the massless eikonal, i.e. Eq.~\eqref{eq:massless-eikonal}, as the shower emission probability. To this end, it's not enough to simply use the massless eikonal and veto emissions below the dead-cone boundary. Instead, we need to integrate the massive eikonal and veto emissions with rapidities below a shifted dead-cone boundary, i.e.
\begin{equation}
	\eta_{\rm cut} = \eta_{\rm dc}
	 + \eta_s\,, \qquad \text{with} \qquad \eta_s = -\frac{1}{2}
\end{equation}
This resembles the phase-space correction in the Lund multiplicity calculation, see Eq.~\eqref{eq:dc-ps-correction}. 

The second ingredient of the construction is the treatment of heavy-quark pair production.
That is, the variable-flavour number scheme imposes that $g\to Q\bar Q$ splittings are only allowed  whenever the transverse momentum of the splitting is larger than the heavy-quark threshold, $k_t > m_Q$.

Operationally, these two ingredients are implemented in the shower as follows.
Emissions are first generated using the standard massless kernels and the usual channel competition.
Once a channel has been selected and a trial emission constructed, a veto is applied depending on the splitting type:
\begin{itemize}
	\item for $Q\to Qg$ splittings, emissions are vetoed if $\eta>\eta_{\rm cut}$,
	\item for $g\to Q\bar Q$ splittings, emissions are vetoed if $k_t<m_Q$.
\end{itemize}
Provided the running coupling is evaluated in the appropriate variable-flavour number scheme and includes the $K_{\rm CMW}$ correction, the resulting integrated emission probability reproduces the logarithmic structure of the massive eikonal.

\begin{figure}
	\centering
	\includegraphics[width=0.4\textwidth,page=2]{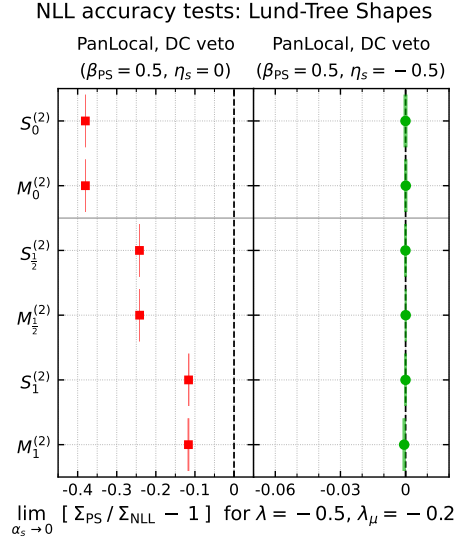}
	\caption{Summary of deviations from NLL for several global observables for $\lambda = -0.5$ and $\lambda_\mu = -0.2$.}
	\label{fig:LTS-simple}
\end{figure}

In Fig.~\ref{fig:LTS-simple} we show NLL logarithmic tests for the LTS observables, where we compare the NLL expectation with the result obtained from the massless \pl shower supplemented with the above veto procedure.
We do so for two variants of the veto, one with $\eta_s = 0$ and the other with $\eta_s = -0.5$. The former does not give the correct NLL result, while the latter does, thus confirming the need for the shift in the dead-cone boundary. 
For this class of observables, mass corrections only enter through the inclusive pattern of soft radiation at NLL accuracy. Thus, the simple shower construction described above is sufficient to reproduce the correct NLL result.

However, we stress that the above procedure does not reproduce the full pattern of correlated soft emissions. In particular, it fails to describe secondary emissions inside the dead-cone region. These emissions contribute at single-logarithmic accuracy for non-global observables and thus the simple shower fails to the energy-in-a-slice test discussed in Sec.~\ref{sec:non-global}.
This means that the above procedure is not viable to achieve NLL-correct showers across all observables. 

\bibliographystyle{JHEP}
\bibliography{MC}

\end{document}